\documentclass[a4paper,twocolumn,11pt]{quantumarticle}
\pdfoutput=1

\makeatletter
\providecommand\@afterenddocumenthook{}
\makeatother

\usepackage[utf8]{inputenc}
\usepackage[english]{babel}
\usepackage[T1]{fontenc}
\usepackage{amsmath,amssymb,amsfonts}
\usepackage{dsfont}
\usepackage{multirow}
\usepackage{makecell}
\usepackage{graphicx}
\usepackage{comment}
\usepackage{enumitem}
\usepackage{placeins}

\usepackage[numbers,sort&compress]{natbib}
\usepackage{hyperref}

\newcommand{\ket}[1]{\left|{#1}\right\rangle}

\newcommand{\braket}[1]{\langle{#1}\rangle}

\newcommand{\ketbra}[2]{\left|{#1}\rangle\!\langle{#2}\right|}

\begin{document}

\title{Clifford Volume and Free Fermion Volume: Complementary Scalable Benchmarks for Quantum Computers}

\author{Attila Portik}
\affiliation{HUN-REN Wigner Research Centre for Physics, 1525 P.O. Box 49, Hungary}
\affiliation{E\"otv\"os Lor\'and University, P.O.\ Box 32, H-1518, Hungary}
\affiliation{Qutility @ Faulhorn Labs, H-1117, Hungary}
\author{Orsolya K\'alm\'an}
\affiliation{HUN-REN Wigner Research Centre for Physics, 1525 P.O. Box 49, Hungary}
\author{Thomas Monz}
\affiliation{University of Innsbruck, Institute of Experimental Physics, Technikerstra\ss e 25/4, 6020 Innsbruck, Austria}
\affiliation{Alpine Quantum Technologies GmbH, Technikerstra\ss e 17/1, 6020 Innsbruck, Austria}
\author{Zolt\'an Zimbor\'as}
\affiliation{HUN-REN Wigner Research Centre for Physics, 1525 P.O. Box 49, Hungary}
\affiliation{QTF Centre of Excellence, Department of Physics, University of Helsinki, Helsinki, Finland}
\affiliation{Algorithmiq Ltd., Helsinki, Finland}

\begin{abstract}
As quantum computing advances toward the late-NISQ and early fault-tolerant eras, scalable and platform-independent benchmarks are essential for quantifying computational capacity in a classically verifiable manner. We introduce two volumetric benchmarks, Clifford Volume and Free Fermion Volume, that assess quantum hardware by testing the execution of random Clifford and free fermion operations.  These two groups of unitaries possess properties that make them ideal for benchmarking: (i) each is individually efficient to simulate classically, enabling verification at scale; (ii) the corresponding gates together form a universal gate set; (iii) they serve as essential algorithmic primitives in practical applications; and (iv) their definitions are formulated abstractly, without explicit reference to hardware-specific features such as qubit connectivity or native gate sets. Together, the proposed benchmarks  enable scalable and fair cross-platform comparisons and track meaningful computational advancement. We demonstrate the practical feasibility of these benchmarks through extensive numerical simulations across realistic noise parameters and through experimental validation on Quantinuum's H2-1 trapped-ion quantum computer, which achieves a Clifford Volume of 34.
\end{abstract}


\section{Introduction}

As quantum computing technologies continue to mature, benchmarks are playing an increasingly central role in evaluating and comparing different hardware platforms~\cite{eisert2020quantum,Proctor2025BenchmarkingQCs,Hashim2025TutorialQCVV, lall2025review}. 
Quantum hardware benchmarks are commonly divided into three categories ~\cite{lorenz2025systematic}: component-level~\cite{Emerson2005, Knill2008, Magesan2012Interleaved,Wallman2014, BlumeKohout2013Tomography, Merkel2013, Nielsen2021GST,noller2025classical}, device-level~\cite{Cross2019,BlumeKohout2020Volumetric,boixo2018characterizing,Erhard2019, HinesProctor2023, demarty2024entropy,Proctor2025Featuremetric, zindorf2025quantum,Zimb2025}, and algorithmic~\cite{Tomesh2022,Chen2022VeriQBench,lubinski2024quantum,dong2021random,martiel2021benchmarking,van_der_Schoot_2024,montanez2025evaluating}. The relative importance of these categories depends on the maturity of the technology, that is, whether one is working with noisy intermediate-scale quantum (NISQ) \cite{Preskill2018NISQ}, late-NISQ \cite{zimboras2025myths}, early fault tolerant \cite{zimboras2025myths, eisert2025mind, katabarwa2024early} or fully fault tolerant devices.
 
Component-level benchmarks probe individual hardware characteristics, such as gate fidelities and qubit coherence times. While such figures of merit remain relevant throughout the evolution of quantum computing, they are particularly important for NISQ devices and for the evaluation of logical components in early fault tolerant systems. Algorithmic benchmarks, in contrast, characterize quantum processing units by measuring how effectively they solve problems via the execution of end-to-end quantum algorithms. 
They are most meaningful for fully fault-tolerant quantum computers, where the reliable implementation of such demanding protocols becomes feasible. In the late-NISQ and early fault-tolerant eras, even though quantum advantage is expected to be attainable~\cite{zimboras2025myths, eisert2025mind,haghshenas2025digital,google2025observation,alam2025fermionic}, practically useful algorithms will likely remain sporadic and heavily platform dependent, often relying on ad hoc methods. Therefore, while monitoring progress in this regime remains important, as demonstrated by the recently launched Quantum Advantage Tracker~\cite{QuantumAdvantageTracker2025}, a different strategy may be needed to quantify computational capacity in ways that allow direct comparisons based on established benchmarking principles \cite{Proctor2025BenchmarkingQCs,lorenz2025systematic, acuaviva2024benchmarking}.
A commonly advocated option is to employ device-level volumetric benchmarks.

Volumetric benchmarks~\cite{Cross2019,BlumeKohout2020Volumetric} assess how reliably a quantum computer executes circuits of increasing width (number of qubits) and depth (number of gate layers). 
Quantum Volume (QV) \cite{Cross2019} is a widely adopted benchmark in this class. The $n$-qubit QV circuit family is specified by the following structure: each circuit is consists of $n$ layers, and every layer is composed of  $n/2$  Haar-random two-qubit gates applied to randomly selected qubit-pairs with no overlaps. Performance evaluation then involves running many random circuit instances of width $n$ on the device and comparing the measured outputs against ideal distributions computed classically.
The example of Quantum Volume highlights three principal challenges confronting volumetric benchmarks: (i) {\it scalability}: classical computation of output probabilities becomes intractable at scale; (ii) {\it algorithmic relevance}: square random circuit structures do not reflect the characteristics of practical quantum algorithms; and (iii) {\it platform independence}: the implicit all-to-all connectivity requirement introduces routing overhead that biases performance metrics toward architectures with dense qubit connectivity. While dense connectivity is in general advantageous, hard-coding such an implicit all-to-all connectivity preference  appears somewhat arbitrary.
Nevertheless, Quantum Volume remains a valuable and historically important benchmark that has provided a common reference point for assessing progress in quantum computing.

While several alternative volumetric benchmarks have been proposed to address these challenges, they typically resolve only a subset of the three identified issues. One benchmark that addresses these issues particularly well is the subcircuit volumetric benchmarking method, which uses subcircuits of varied shapes extracted from target algorithmic circuits \cite{seritan2025benchmarking}. In this work, we take a different approach: rather than examining extracted subcircuits, we consider fully functional algorithmic primitives, i.e., circuits that retain their complete computational function while not constituting complete algorithms.

We introduce two volumetric benchmarks, Clifford Volume (CLV) and Free Fermion Volume (FFV), based on random $n$-qubit unitaries sampled from the Clifford and free fermion groups, respectively. These
$n$-qubit unitaries possess three key properties: they admit efficient classical simulation with compatible input states \cite{aaronson2004,Terhal_2002}, enabling verification at scale; they serve as fundamental algorithmic primitives in practical applications (such as shadow tomography for qubit systems~\cite{huang2020predicting} or fermions~\cite{wan2023matchgate}, as orbital rotations in quantum chemistry~\cite{rubin2022compressing}, or as the free components of many-body physics evolutions~\cite{alam2025fermionic});
and they are defined abstractly, without reference to hardware-specific features like qubit connectivity or native gate sets. 
When these abstractly defined, algorithmically relevant $n$-qubit unitaries are compiled to specific gate sets and connectivity graphs, the resulting gate counts and circuit depths may vary across platforms, as expected. Crucially, however, such architecture-dependent differences emerge naturally through compilation rather than being  built into the circuit definitions. Accordingly, CLV and FFV are explicitly platform-independent at the level of the benchmark task itself. This is intended to reduce architecture-induced bias, where benchmark circuits with a fixed structure could implicitly favour one hardware connectivity or native gate set over another.

This explicit platform-independent design distinguishes our benchmarks from other Clifford-based volumetric benchmarks, which typically employ a randomized benchmarking (RB) framework. The most basic version of Clifford RB applies sequences of Haar-random Clifford unitaries followed by their inverses. However, the substantial circuit depth resulting from many Haar-random Cliffords limits this approach to component-level benchmarking of few-qubit gates \cite{corcoles2013process,mckay2019three}. Making RB scalable requires introducing predefined layers \cite{Erhard2019,Proctor2022MeasuringCapabilities}, 
whose design can introduce topology-dependent bias by implicitly favouring some hardware connectivities over others. Furthermore, a fundamental difference exists between the two approaches: RB methods are designed to be insensitive to measurement errors, as they aim to assess gate and layer fidelities. In contrast, our method deliberately captures both circuit and measurement errors, reflecting our goal to track the joint progress of measurement and circuit quality in a holistic manner.

Importantly, let us point out that the proposed CLV and FFV benchmarks complement each other in several ways. First, they inherently exhibit different sensitivities to connectivity constraints. Second, for fixed width, they probe distinct depth regimes as free fermion circuits typically yield shallower implementations while Clifford circuits require larger depth. Third, Clifford-based benchmarks are often criticized on the grounds of classical simulability, even though this concern appears less relevant in the late-NISQ regime~\cite{merkel2025clifford}. Combining Clifford and free fermion benchmarks however, directly addresses this problem as these families together constitute a universal gate set~\cite{oszmaniec2017universal}, and their composition typically yields classically intractable circuits~\cite{Oszmaniec2022}.
The differences between CLV and FFV sharpen further in the fault-tolerant era, where logical non-Clifford resources dominate the cost budget. While CLV measures only Clifford-level capability, FFV also tests progress in non-Clifford resources (e.g., $T$ gates or Toffolis). The classical simulability of FFV will already become crucial in the early fault-tolerant era. For example, a recent scheme for resource-efficient early fault tolerance~\cite{ibe2025measurement} chart a concrete route to devices that reach a QV of $2^{64}$. However, such QV scores cannot be verified directly, as computing the output probabilities is classically intractable. FFV offers a practical alternative for benchmarking these devices: its T-gate count is comparable to that of QV circuits, yet it remains classically verifiable. Indeed, partly motivated by the use of our FFV benchmark in the fault-tolerant setting, $T$-gate-optimal decompositions of free-fermion operations have recently been developed~\cite{casas2026matchgate}.

The paper is organized as follows. In Sec.~\ref{sec: two_novel_vol_bench} we introduce the theoretical frameworks and determine the benchmark criteria for the CLV (Sec.~\ref{sec: Clifford_Volume_Benchmark}) and the FFV (Sec.~\ref{sec: Free_fermion_Volume}) benchmarks. In Sec.~\ref{sec: Practical_realizations} we discuss details of the practical realization of these benchmarks, and present a numerical analysis of the proposed schemes for a wide range of two-qubit gate errors and readout errors. In Sec.~\ref{sec: Demonstration} we present a demonstration of the CLV benchmark protocol on Quantinuum's H2-1 quantum computer, and determine its CLV score to be $34$. We conclude in Sec.~\ref{sec: Conclusions}.

\FloatBarrier
\section{Two novel volumetric benchmarks} 
\label{sec: two_novel_vol_bench}

\subsection{Criteria for volumetric benchmarks}
\label{sec: criteria_vol_bench}

The first crucial step in defining a volumetric benchmark is to select an appropriate problem set for measuring the capabilities of a quantum processor. As newer generations of quantum devices increased in both qubit count and operation fidelity, they have stepped into the regime where demonstrating quantum advantage becomes feasible, while direct classical verification of their outputs becomes intractable. Consequently, it is essential to design tasks whose results remain classically verifiable, thereby enabling performance comparison in otherwise classically intractable regimes.  To this end, one has to designate a scalable approach,  by allowing the problem size to be dynamically adjusted according to the expected performance of the processor, to pose a meaningful challenge without sacrificing verifiability. 

 We introduce two benchmark protocols tailored for near-term quantum processors, designed to incorporate all previously established properties of volumetric benchmarks while remaining entirely platform-independent. We also provide an open-source software suite that performs all hardware-independent steps of the volumetric benchmarks introduced in this section, including circuit generation in OpenQASM format, enabling platform-specific optimization and execution~\cite{EQCB}.

\subsection{Clifford Volume Benchmark}
\label{sec: Clifford_Volume_Benchmark}

\subsubsection{General considerations}


 In contrast to previously proposed benchmarks, which apply cycles of layers of specific random Clifford gates~\cite{Chen_2023,Merkel2025CliffordProxy}, here we propose to implement randomly chosen $n$-qubit Clifford unitaries as a benchmark task. The stabilizer states generated by the Clifford circuits realizing these random $n$-qubit Clifford unitaries are classically verifiable by measuring the expectation values of $n$-qubit Pauli operators, a process naturally aligned with quantum computing applications. An efficient method for such a verification is Direct Fidelity Estimation (DFE)~\cite{Flammia2011}. In DFE one samples only a constant number of $n$-qubit stabilizers (corresponding to a given additive error and failure probability), measures their expectation values, and combines them into a lower bound on the fidelity. Although DFE provides a practical method, it might still require a significant experimental effort if one wished to use it in an extensive volumetric benchmark task. Therefore, here we choose more 
 simple yet robust criteria to determine whether a quantum processor can adequately implement Clifford unitaries: We require the processor to reliably distinguish stabilizer Pauli strings from non-stabilizer ones. 

The $n$-qubit Clifford group $\mathcal{C}(n)$, is the set of all unitary operators (up to a global phase) that map the $n$-qubit Pauli group to itself under conjugation. 
An $n$-qubit stabilizer state $\ket{\psi}$ is the unique joint eigenstate, with eigenvalue $1$, of $2^n$ commuting $n$-qubit {\it signed Pauli operators} $\{P_l\}_{l=1}^{2^n}$, where each $P_l$ operator, up to a multiplicative factor of $ \pm 1 $,   consists of tensor products of $n$ single-qubit operators (the identity or the single-qubit Pauli matrices $\{I, X, Y, Z\}$). The set $ \left\{P_l\right\}_{l=1}^{2^n} $ forms the stabilizer group $\mathcal{S}$ of the stabilizer state $\ket{\psi}$. The full stabilizer group $\mathcal{S}$ is unique for a given stabilizer state; however, the choice of a generating set $\mathcal{G} \subset \mathcal{S}$ -- a minimal set of $n$ independent operators from which $\mathcal{S}$ can be constructed -- is not unique. One can always choose a set of $n$-qubit Pauli operators that generate the stabilizer group. 
Using the operators from the generating set,  the density operator corresponding to the stabilizer state can be expressed as the projector to the common $+1$ eigenspace of the generators, which  can then be expanded into a sum over the entire stabilizer group
\begin{equation}
\rho \equiv \ketbra{\psi }{\psi} = \frac{1}{2^n} \prod_{P_l \in \mathcal{G}} (I + P_l) = \frac{1}{2^n} \sum_{P_l \in S} P_l.
\end{equation}
From the above definitions, it is straightforward to see that a Clifford unitary transforms a stabilizer state into another stabilizer state
\begin{equation}
\rho'=C \rho\, C^\dagger = C \left( \frac{1}{2^n} \sum_{P_l \in S} P_l \right) C^\dagger = \frac{1}{2^n} \sum_{P'_l \in \mathcal{S'}} P'_l  \, 
\end{equation} 
where $\mathcal{S'} = \left\{P_l' = C P_l C^\dagger | P_l \in \mathcal{S}\right\}_{l=1}^{2^n}$.
Moreover, for any two stabilizer states, there exists a Clifford operation that transforms one into the other. As a direct consequence, any $ n $-qubit stabilizer state $\ket{\psi}_{C}$ can be constructed by applying an appropriate Clifford unitary to the state $ \ket{0}^{\otimes n} $: 
\begin{align}
\rho_C
&= \ketbra{\psi}{\psi}_C
 = C\ketbra{0}{0}^{\otimes n}C^\dagger \\
&= C\left[\frac{1}{2^n}\prod_{l=1}^{n}(I+P_l)\right]C^\dagger
 = \frac{1}{2^n}\prod_{l=1}^{n}(I+CP_lC^\dagger). \notag
\end{align}
where a practical choice for the generator set of the stabilizer group is the set of single-qubit operators $\left\{Z_ 1, \, Z_ 2,  \dots  Z_{n-1}, \, Z_ n \right\}$, where $Z_i$ acts as the Pauli $Z$ operator on the $i$th qubit and as the identity on all other qubits. Since a Clifford unitary maps Pauli operators to Pauli operators up to a phase, each term $ C P_l C^\dagger $ is also a Pauli operator. Thus $\rho_{C}$ is the projector onto the stabilizer state defined by the transformed generators $ \{ C P_l C^\dagger \}_{l=1}^n $.

The $ n $ generators of a given stabilizer group can always be complemented by $n$ additional Pauli operators so that they together form a full generating set for the $ n $-qubit Pauli group, with $ 2n $ linearly independent elements. For any Pauli operator $ P $ in this set, the expectation value of $P$ in state $ \rho $ is given by
\begin{align}
\braket{P} & = \operatorname{Tr} \left(\rho P \right) \\ \notag & = \frac{1}{2^n}\sum_{P_i \in \mathcal{S}}\operatorname{Tr}(P_i\,P)  = 
\begin{cases}
  1,  & P\in\mathcal S,\\
 -1,  & -P\in\mathcal S,\\
  0,  & P,-P\notin\mathcal S.
\end{cases}
\end{align}
since Pauli operators are orthogonal under the trace inner product. Thus, the stabilizer generators have an expectation value of $ 1 $, while the other generators from the complete set have an expectation value of $ 0 $. We will use this fact in our benchmark to decide whether a given quantum computer has successfully implemented a given Clifford unitary, since it is a fundamental demand that the measurement of the expectation values of various Pauli operators can be performed on a quantum computer platform. 

\subsubsection{The Clifford Volume benchmark protocol}

The Clifford Volume benchmark (CLV) is based on the implementation of randomly selected $n$-qubit Clifford unitaries in every step of the protocol until the largest value of $n$ is reached so that the corresponding unitary is successfully realized passing the prescribed criteria, which we detail below. 

In order to reliably determine whether Clifford unitaries of a given size can be implemented correctly, examining only a single specific circuit is insufficient, as this could lead to incorrect conclusions based on isolated cases. On the other hand, the sample size cannot be chosen to be impractically large, as this would unnecessarily complicate the benchmarking protocol and make the evaluation tedious on certain platforms. Therefore, to achieve statistical confidence, we propose that an appropriate number ($ K $) of uniformly sampled  random elements from the $n$-qubit Clifford group be implemented and executed for every value of $n$. For efficient sampling methods of Clifford operations, see Refs. \cite{Berg2020, Bravyi2021}. This approach provides a more accurate assessment of whether the $n$-qubit Clifford operators can be implemented consistently and with high reliability. For each sampled Clifford operation $C^k$ $(k \in [1,K])$, a corresponding quantum circuit realization must be constructed, where the input shall be initialized in quantum state $\ket{0}^{\otimes n}$. These quantum circuits may be optimized in any possible way to achieve better benchmark scores, taking into account factors such as the topology, the native gate set, and other platform-specific parameters (for architecture-aware synthesis of Clifford circuits, see for example \cite{winderl2024architectureaware}). We also allow quantum circuits that approximate the Clifford operation, provided that their results are still validated against the original operators. 

Let us note here that there is no need to initialize the system in a quantum state other than $\ket{0}^{\otimes n}$ as this would, in practice, require the application of yet another $n$-qubit Clifford unitary which can always be combined with the randomly chosen $n$-qubit Clifford unitary that we test. 

In order to evaluate whether the circuit implementation of a given Clifford unitary was successful, we prescribe that for each $\mathcal{C}^k$ Clifford operator,  the  $ n $ generators $\mathcal{S}^k_i$ ($ i \in [1, n] $) of the stabilizer group be assigned, along with $ n $ additional distinct $ n $-qubit Pauli operators $\mathcal{D}^k_i$ ($ i \in [1, n] $) that lie outside the stabilizer group (sometimes called destabilizers), which complete the stabilizer generators to a set of generators for the whole $n$-qubit Pauli group. These operators serve as performance indicators in our protocol, as their expectation values take the value $1$ for stabilizer generators $\mathcal{S}^k_i$ and $0$ for $\mathcal{D}^k_i$, thereby enabling verification of the correct implementation of the Clifford operation.

Let us note that since all the generators $\mathcal{S}^k_i$ commute with each other, one could, in principle, measure their expectation values simultaneously. To achieve this, an $n$-qubit transformation would need to be applied just before the measurements in the $z$-basis. A brief calculation reveals that this operation is equivalent to the inverse of the original Clifford operation. Consequently, this approach would result in a mirror circuit benchmark~\cite{Proctor2022Mirror}, which has the advantage of simplifying the verification of the final result. On the other hand, applying the inverse transformation, would double the depth of the circuit. Moreover, due to its symmetric structure, such a circuit could be insensitive to certain types of systematic errors. For these reasons, we avoid this approach and choose to measure the expectation value of each operator separately.

The total number of operators that could be measured in one step of the benchmark protocol is $2nK$, which scales linearly with $n$. However, for large values of $n$, this can become challenging on NISQ devices. Therefore, in order to keep the benchmark practically feasible, we limit the number of measurements. In each $n$-qubit step of the protocol, we require that the expectation values of $n_{\text{m}} = 4$ different, randomly selected stabilizer and non-stabilizer operators be measured.

On currently available quantum hardware, executing a large number of circuit instances can be particularly demanding. Using only a limited number of Pauli operators per circuit instance therefore significantly reduces the experimental resource requirements of the benchmark, to ensure its practical feasibility. This choice is motivated by the observation that, for a given compiled Clifford circuit, the circuits used to measure different Pauli operators differ only by a final layer of single-qubit basis-change operations right before to measurement in the computational basis. As a result, the dominant contribution to the measured expectation values arises from the quality of the Clifford unitary implementation itself, while the dependence on the specific measured operator is typically limited to minor variations associated with single-qubit rotations and readout. Under these conditions, measuring a small number of randomly selected stabilizer and destabilizer operators provides a practical way to assess the success of a given circuit instance. Moreover, the $\mathcal{S}^k_i$ and $\mathcal{D}^k_i$ operators are randomly selected from the generator set and the expectation values of all elements of this set are known to satisfy the previously discussed properties. In addition, the benchmark is evaluated over an ensemble of randomly sampled Clifford unitaries and repeated measurement rounds. Therefore, even this reduced set of measurements can effectively indicate whether the quantum processor is correctly implementing the intended Clifford unitary.



Upon executing the circuits corresponding to $\mathcal{C}^k$ and carrying out the above mentioned measurements, one obtains a collection of expectation values $\{\braket{S^k_i}\}_{S^k_i \in \mathcal{S}^k}$ and $\{ \braket{D^k_i}\}_{D^k_i \in \mathcal{D}^k} $. Ideally, in an error-free scenario, every $\braket{S^k_i}$ would be equal to $1$, while $\braket{D^k_i}$ would be equal to $0$ up to the statistical uncertainty. However, when errors are present, this does not hold anymore. For each Clifford circuit in the given sample, we estimate the expectation value of the corresponding randomly selected $n$-qubit Pauli operators by measuring them  $L$ times. Each measurement yields one of two possible outcomes, $+1$ or  $-1$, with probabilities due to device noise, imperfections, or other system-dependent effects. We deliberately avoid using the average of measured expectation values as a performance indicator because we aim to detect scenarios where a device 
exhibits significantly poor performance in certain individual cases. Using the average alone could hide such anomalies; thus, we use the minimum value from the entire set instead. This stricter approach enables us to monitor whether quantum devices can reliably distinguish between expectation values that should ideally be equal to $0$ and $1$.


In order to verify that the two measured distributions are sufficiently distinguishable, we impose that for every Clifford circuit the measured expectation value of each stabilizer operator exceeds a predefined threshold value $\tau_S$. Analogously, for the non-stabilizer operators, the ideal expectation value is zero; however, due to errors, deviations in either direction may occur. We therefore require that the absolute value of the measured expectation values of the non-stabilizer operators remains below a threshold $\tau_D$. In practice, expectation values are estimated from a finite number of measurement shots and hence are subject to statistical uncertainty. In order to account for this, we extend the success criteria by explicitly incorporating the corresponding statistical uncertainties. Specifically, we require that the stabilizer expectation values exceed $\tau_S$ by at least two standard deviations, and that the absolute value of the non-stabilizer expectation values remains below $\tau_D$ by at least two standard deviations. This ensures that the benchmark thresholds are not violated within the estimated statistical uncertainty and prevents marginal statistical fluctuations from being misidentified as failures. Motivated by the use of a two-standard-deviation confidence margin in Quantum Volume, we adopt the same margin for these single-instance acceptance criteria. Formally, these conditions can be expressed as
\begin{equation}
\begin{cases}
\displaystyle \braket{\mathcal{S}_i^{k}}  - 2\sigma_{\mathcal{S}^k_i} \ge \tau_S, \\[3mm]
\displaystyle \left| \braket{\mathcal{D}_i^{k}} \right| + 2\sigma_{\mathcal{D}^k_i} \le \tau_D,
\end{cases}
\qquad \forall \, i,k
\end{equation}
where the standard deviation of the measured expectation value is
\begin{align}
\sigma_{\mathcal{P}_i^{k}} = \sqrt{\frac{1-\langle \mathcal{P}_i^{k}\rangle^2}{L}}, \label{sigma_S_D}
\end{align}
with $\mathcal{P}=\left\lbrace \mathcal{S},\mathcal{D}\right\rbrace$.

In addition to the worst-case conditions defined above, we also impose a criterion on the average performance. Specifically, for every sampled Clifford unitary, we require that the average stabilizer expectation value lies at least five standard deviations above the stabilizer threshold, and that the average destabilizer expectation value lies at least five standard deviations below the destabilizer threshold. Formally, these conditions read
\begin{equation}
\begin{cases}
\displaystyle
\overline{\braket{\mathcal{S}^{\,k}}} - 5\,\overline{\sigma}_{\mathcal{S}^{\,k}} \ge \tau_S, \\[3mm]
\displaystyle
\left| \overline{\braket{\mathcal{D}^{\,k}}} \right| + 5\,\overline{\sigma}_{\mathcal{D}^{\,k}} \le \tau_D ,
\end{cases}
\qquad \forall\, k .
\end{equation}
where $\overline{\braket{\mathcal{S}^{\,k}}} $ and $\overline{\braket{\mathcal{D}^{\,k}}} $ denote the averages of the measured stabilizer and destabilizer expectation values corresponding to the $k$th Clifford unitary, and $\overline{\sigma}_{S^k}$ and $\overline{\sigma}_{D^k}$ are the associated standard deviations of these averages, defined as 
\begin{align}
\overline{\sigma}_{\mathcal{P}^k} =\frac{1}{4}\sqrt{\sum_{i=1}^{4} \sigma^{2}_{\mathcal{P}^{k}_{i}} }\,.
\label{sigma_S_D_avg}
\end{align}
The five-standard-deviation margin is chosen to make the average-performance criterion comparable to, but stricter than, the two-standard-deviation single-operator criterion. Since averaging reduces the propagated uncertainty, a larger numerical prefactor is used.

This average-performance criterion together with the worst-case criterion ensures that each sampled Clifford unitary exhibits statistically significant and robust behavior, rather than allowing good performance to arise only after averaging over multiple Clifford instances. Our choice of two and five standard deviations reflects a practical trade-off between statistical confidence and experimental resource constraints and is consistent with common practice in experimental benchmarking.

Given the significant variability in quality and computational power among current quantum processors, imposing overly strict benchmark conditions is impractical. Therefore, when selecting the threshold value $\tau_D$, it is crucial to account for the specific properties of the measured operators. Although setting this parameter is inherently somewhat arbitrary -- and there are more sophisticated statistical approaches available -- we prefer a straightforward solution. We choose a moderately low empirically informed threshold that captures the minimum acceptable performance level expected from current quantum hardware: For stabilizer operators we set the threshold to $\tau_S = \frac{1}{e}$. This choice is motivated by the fact that, under a baseline error model, such as a global depolarizing channel with a fixed error probability per circuit layer, the expectation values decay roughly exponentially with circuit depth, thus, in the optimal case, with the number of qubits. In such cases, reaching $1/e$ ($\sim37\%)$ of the initial expectation value is a natural way to define the effective "lifetime" of the system. 
This also aligns well with thresholds commonly adopted in existing quantum benchmarking methods~\cite{Proctor2022MeasuringCapabilities}.

In contrast, the measurement of non‑stabilizer operators ideally yields an expectation value of zero, because in the noise‑free case, measurements equally produce $+1$ and $-1$ outcomes. Errors may affect this average, but since the ideal value is already zero, small deviations cannot be diagnosed as a decay. Instead, noise can lead to the appearance of expectation values lying further away from zero, which justifies using a specific threshold for these expectation values too. We therefore choose a threshold $\tau_D = \frac{1}{2e}$ that is moderately close to the ideal value, yet above the expected level of statistical fluctuations. In a noise‑free scenario, such a large deviation would be extraordinarily unlikely. Thus, observing an expectation value above this threshold in experiments indicates -- with extremely high confidence -- that noise is present.

Based on the above detailed considerations and criteria we describe the steps for the practical implementation of the CLV benchmark protocol in Appendix~\ref{appendix : CV_step_by_step}. 
The benchmark does not prescribe a unique search algorithm for selecting the candidate values of $n$. Rather, the choice is left to the user, who may use prior calibration data, emulator results, preliminary trials, or device-specific error models to allocate experimental resources efficiently. This platform-dependent search procedure only determines which candidate qubit counts are tested in detail, it does not effect the definition of the CLV score.




\subsection{Free-fermion Volume}
\label{sec: Free_fermion_Volume}

We propose the Free-Fermion Volume (FFV) benchmark as a counterpart to the Clifford Volume benchmark. Although they share many similarities, Clifford and free-fermion unitaries span distinctly different regions of the full unitary group. More specifically, in the FFV benchmark, we assess how effectively a quantum computing platform can capture and represent the abstract structure of the special orthogonal group, as detailed below.

\subsubsection{General considerations}

The Hilbert space of a system of $d$ fermionic modes is described by the Fock space, constructed as the direct sum of anti-symmetric subspaces:
\begin{equation}
\mathcal{H}_\text{Fock}(\mathbb{C}^d) = \bigoplus_{n=0}^d \wedge^n (\mathbb{C}^d),
\end{equation}
where $\wedge^n (\mathbb{C}^d)$ is the subspace corresponding to states with exactly $n$ fermions. The fermionic creation ($f_j^\dagger$) and annihilation ($f_j$) operators act naturally on this space, and they satisfy the canonical anti-commutation relations:
\begin{equation}
\{f_j, f_k^\dagger\} = \delta_{j,k}, \quad \{f_j, f_k\} = \{f_j^\dagger, f_k^\dagger\} = 0.
\end{equation}
The vacuum state, denoted $|0_F\rangle$, is the state with no particles, and the creation operators $f_j^\dagger$ construct Fock basis states:
\begin{equation}
|x\rangle = (f_1^\dagger)^{x_1}(f_2^\dagger)^{x_2} \cdots (f_d^\dagger)^{x_d} |0_F\rangle,
\end{equation}
where $x = (x_1, x_2, \ldots, x_d) \in \{0, 1\}^d$ encodes the occupation numbers of the fermionic modes.

It is  also useful to introduce the Majorana operators $m_j$. These operators are linear combinations of creation and annihilation operators defined as
\begin{equation}
m_{2j-1} = f_j + f_j^\dagger, \quad m_{2j} = -i(f_j - f_j^\dagger ),
\end{equation}
with anti-commutation relations $\{m_j, m_k\} = 2\delta_{j,k}$.
Majorana operators provide a convenient basis for analyzing fermionic systems due to their self-adjoint property and direct connection to physical observables.

Free fermionic evolutions are fundamental  in quantum many-body physics. They are defined by their generator, which is a Hamiltonian that is quadratic in terms of Majorana operators
\begin{equation}
H = \frac{i}{4} \sum_{j,k} A_{j,k} m_j m_k \, ,
\end{equation}
where $A$ is a real antisymmetric matrix ($A = -A^\mathsf{T}$).
The unitary $U=e^{-iH}$ generated by such a Hamiltonian acts on the subspace of single Majorana operators in an especially simple manner
\begin{equation}
U^\dagger\,m_j\,U
= \sum_k O_{jk}\,m_k,
\end{equation}
where  $O = e^{A} \in SO(2L)$ is an orthogonal matrix with determinant 1~\cite{Terhal_2002, zimboras2014dynamic}.

The Jordan-Wigner transformation \cite{jordan1928paulische} establishes an equivalence between fermionic and qubit systems, enabling the representation of fermionic systems on qubits. It is most naturally defined as a unitary $\left(\mathcal{V}_\text{JW}:  \mathcal{H}_\text{Fock}(\mathbb{C}^d) \to (\mathbb{C}^2)^{\otimes d}  \right)$ that simply maps Fock basis states to computational basis states \cite{miller2023bonsai}:
\begin{equation}
\mathcal{V}_\text{JW} \left( (f_1^\dagger)^{x_1}(f_2^\dagger)^{x_2} \cdots (f_d^\dagger)^{x_d} |0_F\rangle \right) = \bigotimes_{p=1}^d |x_p\rangle
\end{equation}
for all $(x_1, x_2, \ldots, x_d) \in \{0, 1\}^d$. On the operator level, the adjoint action of $\mathcal{V}_\text{JW}$ maps fermionic operators to Pauli operators on the qubit system as follows:
\begin{align}
m_{2p-1} & \mapsto Z_1 Z_2 \cdots Z_{p-1} X_p, \\ \notag  m_{2p} & \mapsto Z_1 Z_2 \cdots Z_{p-1} Y_p,
\end{align}
where $X_p, Y_p, Z_p$ are the Pauli operators acting on the $p$th qubit, and $Z_1 Z_2 \cdots Z_{p-1}$ introduces the non-locality inherent to fermionic statistics.

\subsubsection{The Free-Fermion Volume benchmark protocol} 

In the FFV benchmark, we focus on the implementation of randomly chosen $n$-qubit free-fermion unitaries, which correspond to matrices $O \in SO(2n)$ and thereby form a suitable problem set for volumetric benchmarking.

The FFV benchmark protocol consists of three main steps: (i) prepare an appropriate initial state, (ii) evolve this initial state using a randomly chosen unitary operator from the free-fermion problem set and (iii) measure carefully chosen expectation values to assess whether the evolution has been accurately executed. By repeating these steps across multiple rounds of experiments and analyzing the resulting expectation values, we can derive an overall benchmark score that quantifies the performance of the quantum processor.

In order to perform the FFV benchmark, we generate a random sample of $n$-qubit free-fermionic operations by uniformly sampling from the group of special orthogonal matrices. For each randomly chosen element $O \in SO(2n)$, the corresponding unitary operation is defined via the free-fermionic time evolution:
\begin{equation}
F(t) = \exp \left( \frac{1}{4} \sum_{i,j=1}^{2d} [\log(O(t))]_{ij} m_i m_j \right) \,.
\end{equation}

In order to evaluate the benchmark, for each unitary in the sample, we prepare the system in a randomly chosen uniform superposition of fermionic states, i.e., for a randomly chosen $i \in \left[1,2n \right]$ index, we prepare the state $m_i\ket{0}^{\otimes n}$. Then, considering the time evolution governed by the free-fermion operator, we can verify the orthogonality of the assigned $O$ matrix by measuring the expectation values of the corresponding Majorana mode operators in the quantum state generated by the free-fermion unitary, given by
\begin{align}
\braket{m_j}_{F \rho_i  F^\dagger} &=  
\operatorname{Tr}\left(\rho_i m_j(t)\right) \\ \notag &= 
\operatorname{Tr}\left(F \rho_i  F^\dagger \, m_j   \right) \, ,
\end{align}
where we have omitted notation for time dependence $t$ for the sake of simplicity, since we only consider a single time step.
Because the time evolution is free fermionic, we can describe it directly using the operator $O$, so that
\begin{align}
\braket{m_j}_{F \rho_i  F^\dagger} &= \operatorname{Tr}\left( \rho_i \sum_k O_{jk}\, m_k  \right) \\ \notag &= \sum_k O_{jk} \braket{m_k}_{\rho_i}\, ,
\end{align}
where we used the linearity of the trace operation $\braket{m_j}_{\rho_i} = \sum_k O_{j,k} \operatorname{Tr}\left(\rho_i m_k \right).$ Considering the expectation value of the original Majorana operators in the initial state $\ket{i}$
\begin{equation}
\braket{m_j}_{\rho_i} = \operatorname{Tr} \left( \rho_i m_j\right) = \delta_{ij} \, ,
\end{equation}
we can identify the expectation value of the transformed Majorana operator as 
\begin{equation}
\braket{m_j}_{F \rho_i  F^\dagger} = \sum_k O_{jk} \delta_{ik} = O_{ji} \, .
\label{eq : expected}
\end{equation}
This simple equation makes it possible to test the orthogonality relation by measuring the the expectation value of the Majorana operators,
\begin{equation}
\sum_k O_{kj} \braket{m_k}_{F \rho_i  F^\dagger} = \sum_k O_{ki} O_{kj} = \delta_{ij} \, .
\end{equation}

The quantum device being tested passes our benchmark for a certain number of qubits $n$ if it proves that it is able to correctly represent the structure of the $SO(2n)$ group through the orthogonality relation. In practice this means that we can measure accurately enough the expectation values of the Majorana operators to be able to reconstruct the combinations in the relation and to reliability differentiate between the values equal to $0$ and $1$ in the ideal case.

Let us note that, similarly to the Clifford Volume benchmark, the number of required expectation values grows linearly with the number of qubits. In particular, the total number of measurements for $K$ randomly selected free-fermion operators is $ 2nK$. Despite this linear scaling, the benchmark could remain infeasible on current platforms, since for example, for 100 qubits it would require $200K$ circuit executions. Therefore, in order to keep the benchmark practical, we introduce a similar limitation on the number of measurements as in the case of the CLV benchmark. Specifically, for an $n$-qubit system we prescribe that only
\begin{equation}
N(n) =
\begin{cases}
2n, & \text{for } n \leq 10, \\[2mm]
20 +  \left\lfloor\frac{n}{5} \right\rfloor, & \text{for } n > 10
\end{cases}
\end{equation}
different Majorana operators be measured. We must point out that here one cannot simply select a random finite fraction of the operators as was done in the case of the Clifford Volume benchmark. This is because the aim is to evaluate a linear combination of measured expectation values, each of which may be individually close to zero. Consequently, one needs to carefully select a suitable subset of operators for the measurements to ensure that the results remain meaningful and capable of validating the implementation of the unitary operation. Moreover, in this case, the cut-off imposes more severe constraints on the measured quantities. If we choose a certain fraction $r$ of the operators randomly, the value of the evaluated linear combination decreases significantly. In order to see this, let us consider the case where we randomly select $r = 10\%$ of the matrix entries. A row of $O$ has $2n$ entries; when choosing a $10\%$ fraction randomly, the expected sum of the squares of these selected entries will be given as
\begin{equation}
\left\langle \sum_{i=1}^{\lfloor 2n\,r \rfloor} (O_{ij})^2 \right\rangle
\approx \sum_{i=1}^{\lfloor 2n\,r \rfloor} \frac{1}{2n}
\approx
\frac{\lfloor 2n\,r \rfloor}{2n}
\approx
r.
\end{equation} 
Even on an ideal, noise-free quantum device, distinguishing $0.1$ from $0$ would require a large number of shots, making this approach challenging. Therefore, in order to be able to examine the effects of noise, we propose to select a fraction $r$ of the entries with the largest absolute value (or in other words, we omit the elements that contribute least to the overall value). Since these entries contribute the most to the sum of squares, this approach produces a significantly larger value than random selection. For the $r = 10\%$ case this leads to the value
\begin{equation}
\left\langle \sum_{i\in J} (O_{ii})^2 \right\rangle \approx 0.439,
\end{equation}
where $J$ denotes the indices of the chosen elements (for details see Appendix~\ref{appendix : reduce}). In addition, since the value of the reduced linear combination can be efficiently computed based on the selected matrix elements, we employ the reduced and renormalized linear combination

\begin{align}
R_{ij}^{(J)}
\equiv
\begin{cases}
\dfrac{\left\langle \sum_{k \in J} O_{ki} \braket{m_k}_{\rho_i} \right\rangle}
{\left\langle \sum_{i\in J} (O_{ij})^2 \right\rangle }
~\text{if}~ j=i\\[5mm]
\left\langle \sum_{k \in J} O_{kj} \braket{m_k}_{\rho_i} \right\rangle
~\text{if}~ j\neq i
\end{cases}
\label{reduced_lincomb}
\end{align}

for the benchmark, rather than the full linear combination, in order to reduce the resource demands of the procedure. The quantity defined in Eq.~(\ref{reduced_lincomb}) performs significantly better than its version based on randomly selected matrix elements, as it spans the same range as the full linear combination, but here each term can be measured more accurately due to being significantly larger than zero. 

In order to define a success criterion for  the FFV benchmark, we propose a condition analogous to the one used in the CLV benchmark. While the Clifford Volume Benchmark relies on stabilizer and non-stabilizer operators, here we use two distinct linear combinations of the measured expectation values. These combinations correspond essentially to two rows of the $O$ matrix: the first linear combination is defined by the row of $O$ corresponding to the initial state, which should yield a value close to 1, since each row is normalized ("parallel" linear combination), while the second is defined by a randomly chosen row, which should yield a value close to 0 due to the orthogonality of different rows (”orthogonal” combination).

We impose the condition that each measured value corresponding to the "parallel" linear combination must exceed a specified threshold, while the absolute values of the measurements for the "orthogonal" combinations must remain below a given threshold up to the statistical uncertainty. Following the same reasoning as before, we set these threshold values to $\frac{1}{e}$ for the parallel case and $\frac{1}{2e}$ for the orthogonal case. Formally, using the quantity defined in Eq.~\eqref{reduced_lincomb},
this can be expressed as
\begin{equation}
\begin{cases}
\displaystyle
R_{ii}^{(J)}
- 2\,\sigma_{\parallel}^k
\;\ge\;
\dfrac{1}{e},
\\[3mm]
\displaystyle
\left|
R_{ij}^{(J)}
\right|
+ 2\,\sigma_{\perp}^k
\;\le\;
\dfrac{1}{2e},
\end{cases}
\,
\forall\, k .
\label{FFV_criteria}
\end{equation}
The statistical uncertainties $\sigma_{\parallel}^k$ and $\sigma_{\perp}^k$
are obtained by propagating the shot-noise uncertainties of the individual
Majorana expectation values included in the linear combinations
\begin{align}
\label{eq : sigmas}
\sigma_{\parallel}^k
&=
\frac{\sqrt{
\sum_{\ell\in J_k}
(O_{\ell i}^k)^2
\frac{1-\braket{m_l}_{F\rho_iF^\dagger}^2}{L}
}}{\sum_{\ell\in J_k}(O_{\ell i}^k)^2},
\\[3mm] \notag
\sigma_{\perp}^k
&=
\sqrt{
\sum_{\ell\in J_k}
(O_{\ell j}^k)^2
\frac{1-\braket{m_l}_{F\rho_iF^\dagger}^2}{L}
}.
\end{align}

The FFV benchmark protocol shall begin with circuits for $ n = 1 $ qubit. If the  procedure produces results corresponding to the success criteria, then the number of qubits involved in the next iteration of the benchmark protocol is increased by one. In line with the core idea of our benchmark framework, we deliberately do not prescribe an optimized search pattern or impose strict rules on how the $n$-qubit subset must be selected from the available qubits. Instead, we grant the device provider the flexibility to choose the ideal subset of qubits at each step, as our primary interest is in assessing the peak performance of the device. Similarly, we do not mandate a specific search strategy for determining the maximum qubit count. Any suitable search algorithm may be applied to accelerate and streamline the process. However, a qubit number $ n_{\max} $ can only be considered valid if the device can demonstrably exceed the threshold for all $ n \leq n_{\max} $. Although we offer the flexibility to use any search algorithm and selection pattern, it is worth noting that, given the current generation of devices, the number of qubits passing the test is unlikely to exceed a few tens. As a result, finding the maximal value does not pose a significant challenge. Following the considerations and criteria discussed above, the practical implementation steps of the Free-Fermion Volume benchmark protocol are presented in Appendix~\ref{FFV_step_by_step}.

\FloatBarrier
\section{Practical realizations of the benchmark protocols}
\label{sec: Practical_realizations}

Although one of the key strengths of our framework lies in the flexibility it offers users to optimize their implementations without having to adapt to rigid constraints, we find it useful to provide  baseline implementations for both of the proposed benchmarks. In this section, we present straightforward implementations by decomposing Clifford and free fermion operations into standard single- and two-qubit gates. We evaluate our benchmarks using a simple virtual backend that assumes all-to-all qubit connectivity, i.e., no constraints on the coupling map. As we have pointed out above, this assumption is not required by the benchmark. We use it because it significantly simplifies the compilation of two-qubit gates, as it avoids the need for routing or the addition of SWAP operations which are often required on real hardware due to limited connectivity. While this is not characteristic of all physical devices, it provides us the most direct path from theoretical design to circuit implementation. It is important to note that we do not employ optimized synthesis or compilation techniques. Instead, we use a brute-force approach, resulting in deeper circuits than would be achievable with an ideal compilation.

The virtual backend that we use is equipped with a simple noise model. Rather than simulating the intricate error characteristics of a specific quantum device, we adopt a general, easily customizable noise model designed to capture the most important sources of noise: We assume that two-qubit gates are the dominant source of error together with readout errors, while single-qubit gates are assumed to be ideal. We model two-qubit noise by using depolarizing channels after each two-qubit gate, characterized by a probability $ p_{2Q} $, and we simulate readout errors as independent bit-flip errors on each qubit immediately before measurement, with probability $ p_M $.

The purpose of this implementation is not to provide a fully optimized, platform-specific benchmarking process, but rather to demonstrate the general applicability and robustness of our benchmarking protocols under a well-defined, consistent noise model.

\subsection{Realization of the Clifford Volume benchmark}

In order to demonstrate how to implement the Clifford Volume benchmark protocol we utilize the STIM simulator \cite{gidney2021stim}, a specialized stabilizer circuit simulator capable of efficiently handling Clifford operations. This capability makes it particularly well-suited for simulating large-scale CLV benchmark circuits. Evaluating the benchmark protocol on such simulated virtual hardware allows us to demonstrate the protocol under idealized conditions as well as some basic controlled error models. For the emulated CLV benchmark, we assume that the virtual backend is capable of directly implementing the generating gate set $\{ H, S, \text{CNOT} \}$, as well as single-qubit measurements in the computational $Z$-basis.

A key technical challenge in implementing the CLV benchmark lies in efficiently decomposing arbitrary Clifford unitaries into circuits comprising only elementary Clifford gates. For this purpose, we employ the decomposition procedure implemented in STIM, that is based on Gaussian elimination. Instead of explicitly constructing and manipulating large $2^n \times 2^n$ unitary matrices, this method operates directly on the more compact, $2n \times (2n+1)$ tableau representations of Clifford unitaries. By systematically cancelling off-diagonal elements in the tableau, it transforms an arbitrary Clifford tableau into its canonical form. Each step of this process corresponds to applying one or more elementary Clifford gates from the set $\{H, S, \text{CNOT}\}$. Consequently, this procedure generates a quantum circuit with $O(n^2)$ gates and a depth that scales as $O(n^2)$. For additional details regarding the underlying decomposition methodology, see Refs.~\cite{aaronson2004} and ~\cite{gidney2021stim,vandennest_2008}.

\begin{figure}[t]
\includegraphics[width=\linewidth]{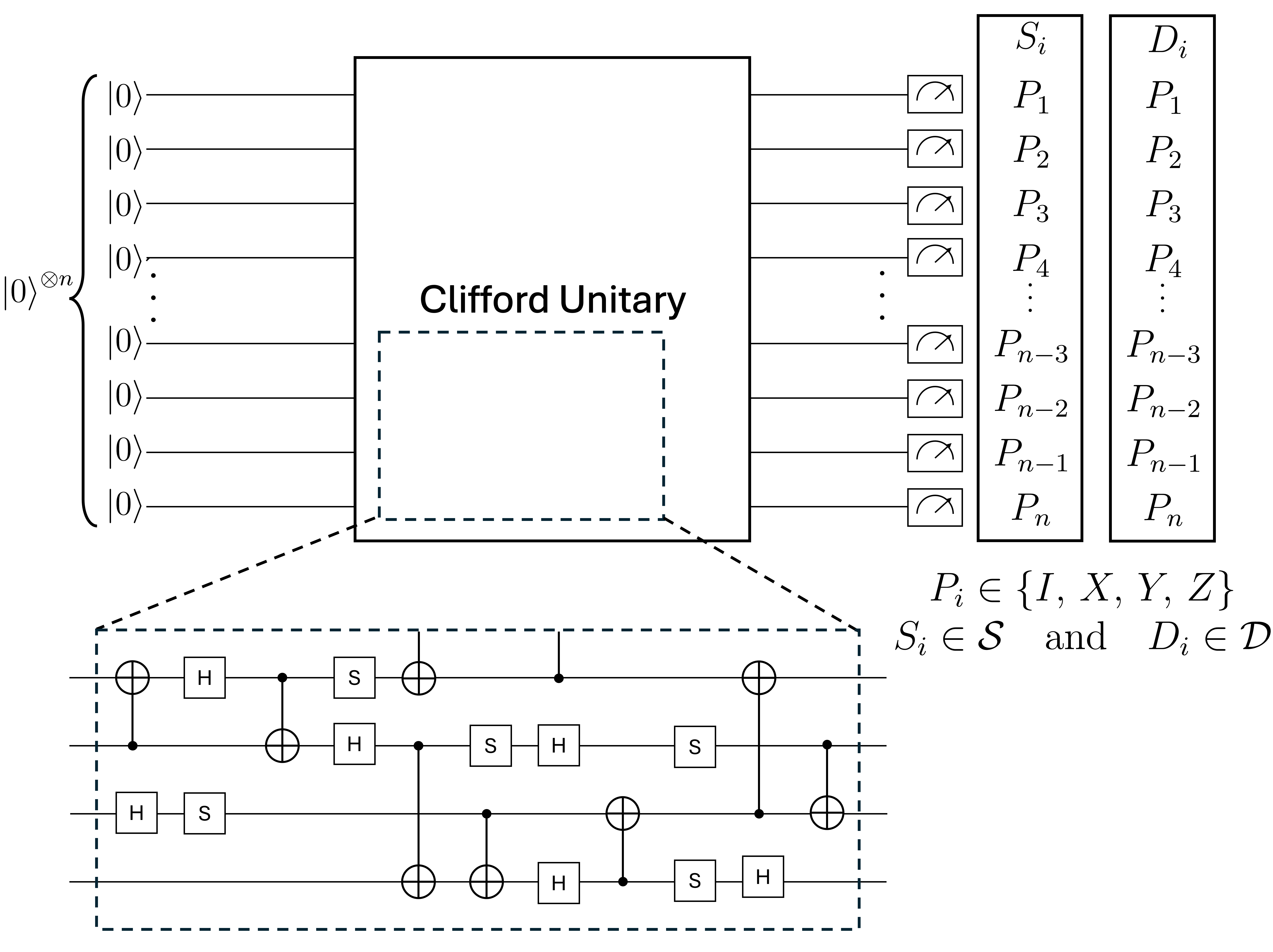}
\caption{\label{fig : Clifford-Volume-Benchmark-scheme}Graphical representation of the $n$-qubit step in the Clifford Volume benchmark protocol. The quantum circuit is initialized in the $\ket{0}^{\otimes n}$ state. Then, a Clifford unitary is applied, which is decomposed and compiled according to the characteristics of the target platform. The circuit is run for a prescribed number of times (see Appendix~\ref{appendix : CV_step_by_step}) in order to evaluate the expectation values of selected $n$-qubit Pauli operators $P_i \in \{I, X, Y, Z\}$, grouped into subsets $S_i \in \mathcal{S}$ and $D_i \in \mathcal{D}$, corresponding to stabilizer and destabilizer sets, respectively.}
\end{figure}

The CLV benchmark protocol requires measuring the expectation values of the chosen $n$-qubit Pauli operators. Current hardware is often restricted to computational ($Z$-basis) measurements, so to accommodate this, we append a layer of single-qubit Clifford gates implementing the required basis changes: $H$ to  measure $X$,  $S^\dagger$ followed by an $H$ to measure $Y$, and no additional gates when measuring $Z$. See Fig.~\ref{fig : Clifford-Volume-Benchmark-scheme} for a schematic depiction of the CLV protocol.

\begin{figure}[t]
\centering
\includegraphics[width=\linewidth]{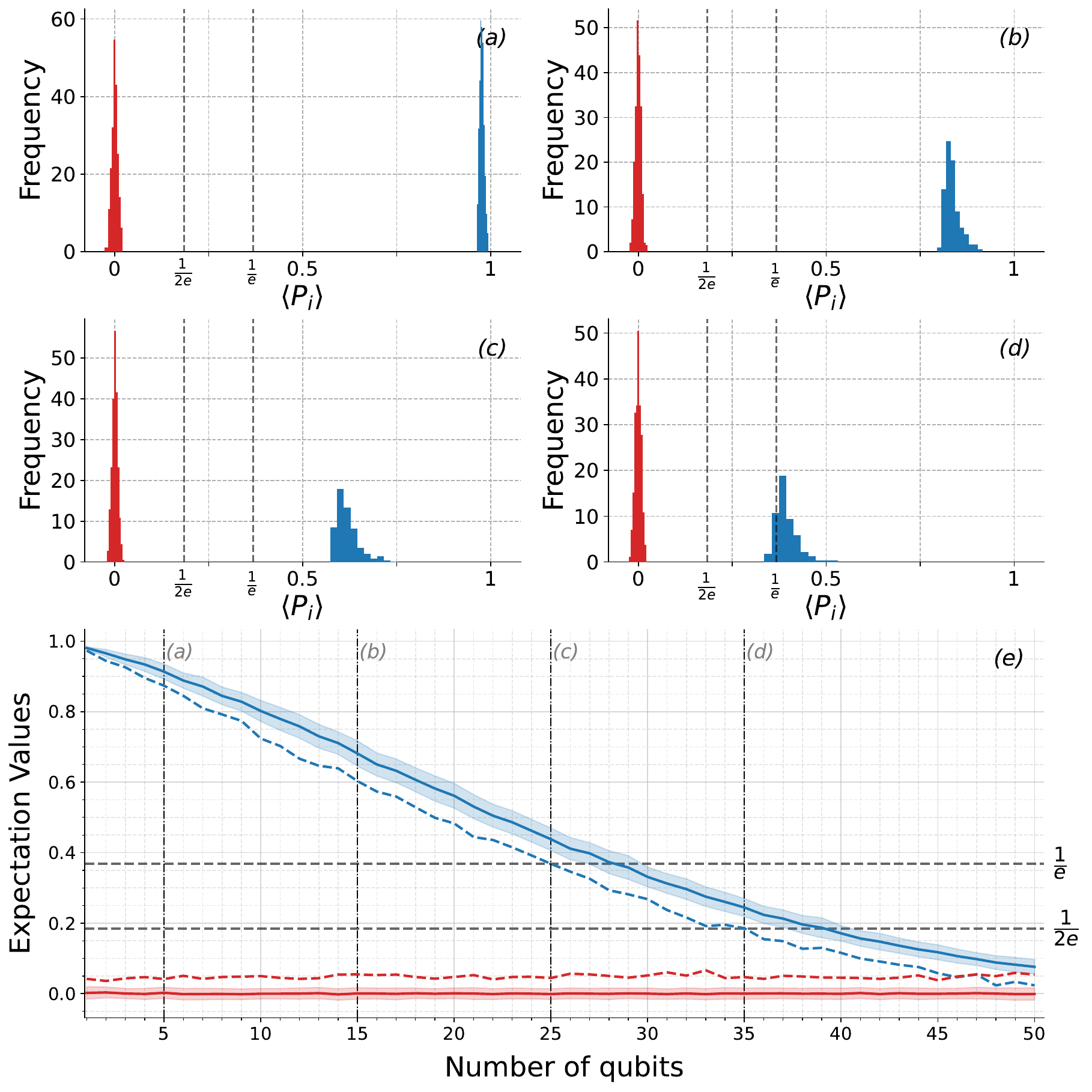}
\caption{\label{fig : Clifford_selected}
Distribution of simulated expectation values for randomly selected Pauli operators in the CLV benchmark, with readout error $p_m = 10^{-2}$ and two-qubit gate error $p_{2Q} = 10^{-3}$. Subfigures (a)–(d) show histograms of the expectation values $\langle P_i \rangle$ for stabilizer (blue) and destabilizer (red) operators at different qubit counts: $n = 5$, $15$, $25$, and $35$, respectively. The horizontal dashed lines indicate the  threshold values: $\frac{1}{e}$ for stabilizers and $\frac{1}{2e}$ for destabilizers. Subfigure (e) shows the average expectation values of stabilizers (blue) and destabilizers (red) as a function of the number of qubits. Shaded regions indicate the standard deviation over the ensemble, while dashed lines represent the maximum deviation from the ideal value for each ensemble. Vertical dashed lines correspond to the qubit counts shown in subfigures (a)–(d).
}
\end{figure}

For our simulated scenarios, we defined a range of error parameters that includes realistic values—spanning from worst-case noise levels to those characteristic of near fault-tolerant devices—in order to cover the full spectrum of quantum hardware expected in the NISQ and late-NISQ eras. Specifically, we explored two-qubit gate error and measurement error probabilities ($p_{2Q}$ and $p_m$) ranging from $10^{-2}$ to $5\cdot10^{-5}$. Figure~\ref{fig : Clifford_selected} shows an example of how the average expectation values of randomly selected stabilizer and destabilizer operators vary as a function of the number of qubits for the noise parameters $\left(p_{2q}, p_{2m}\right) = \left(10^{-3}, 10^{-2}\right)$. These data characterize the general behavior of stabilizer expectation values under noise. For each qubit number $n$, we sample ten random $n$-qubit Clifford unitaries, and for each Clifford instance we evaluate the expectation values of $\max\{10,n\}$ randomly selected stabilizer and destabilizer operators within the assumed noise model. As $n$ increases, the expectation values of the stabilizer operators exhibit a gradual decay and eventually approach the predefined threshold. In contrast, the expectation values of the destabilizer operators remain approximately constant on average. At the same time, the standard deviation and extreme values of the measured distributions -- which are central to our benchmark criteria -- indicate a growing spread in the distribution as the number of qubits increases, progressively approaching the corresponding threshold values.

It is worth noting that the condition for the destabilizer values is generally easier to satisfy than that for the stabilizers. This is because noise tends to reduce expectation values towards zero, and the error-free expectation values of destabilizers already fluctuate around zero. Nevertheless, as our criterion considers the maximum deviation within the measured values, it remains sensitive to changes in these values as well.

\begin{figure}[t]
    \centering
    \includegraphics[width=\linewidth]{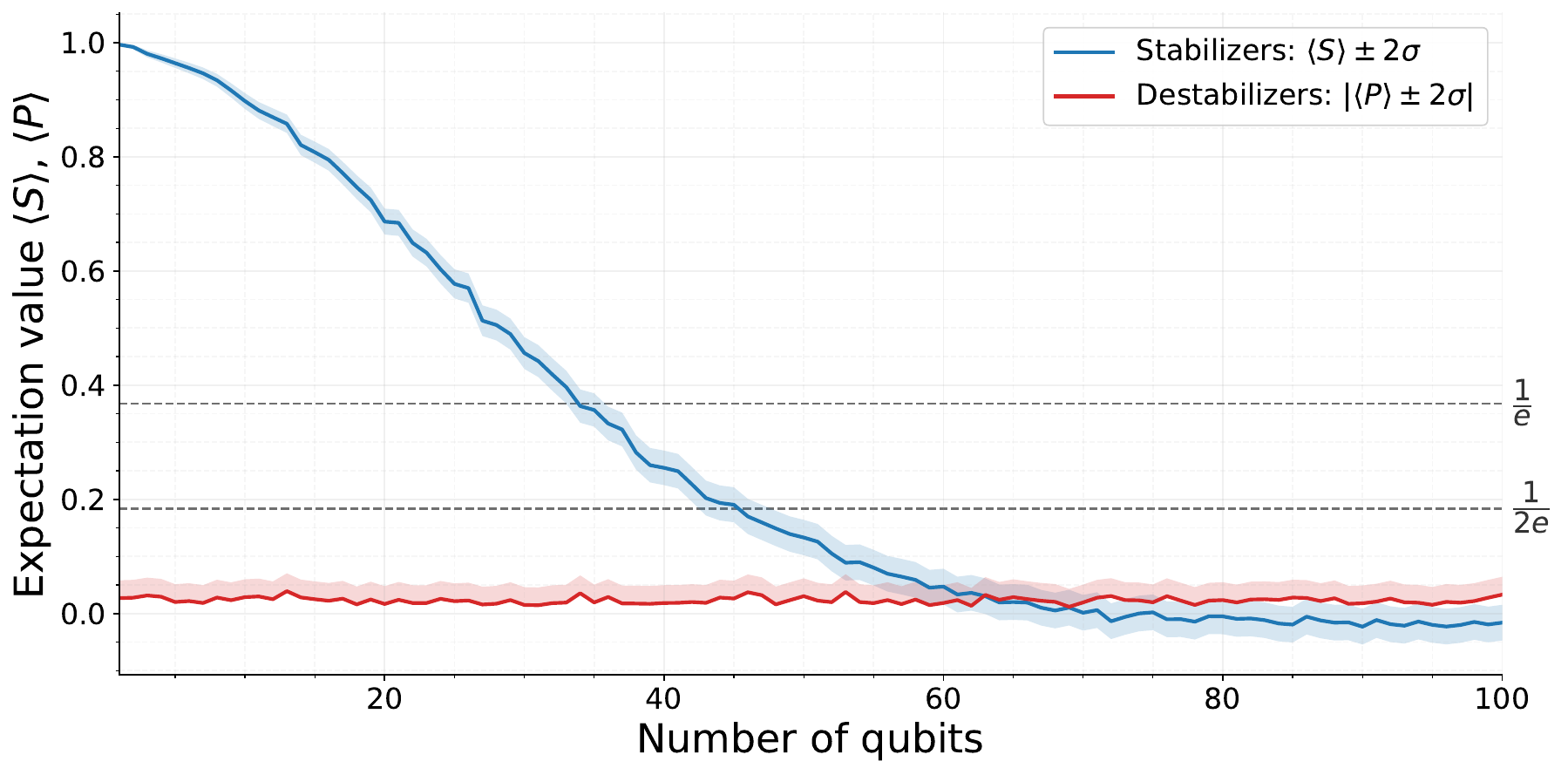}
    \caption{
    The simulated minimum stabilizer expectation values and the maximum destabilizer absolute  expectation values as a function of the number of qubits in the CLV benchmark . Expectation values are estimated by simulating $4096$ measurement shots per observable. For each qubit number $n$, four randomly sampled $n$-qubit Clifford unitaries are implemented. The blue curve shows the smallest measured expectation value among all sampled stabilizer operators, while the red curve shows the largest absolute value of the measured expectation values of destabilizer operators (i.e., the one farthest from zero). Shaded regions indicate the $\pm 2\sigma$ statistical uncertainty used in the benchmark threshold criteria. Horizontal dashed lines mark the acceptance thresholds $\tfrac{1}{e}$ for stabilizers and $\tfrac{1}{2e}$ for destabilizers. 
    }
    \label{fig:Clifford_selected}
\end{figure}

Figure~\ref{fig:Clifford_selected} shows the benchmark results obtained when the
Clifford Volume benchmark protocol itself is evaluated. For each qubit number $n$, four randomly chosen $n$-qubit Clifford unitaries are simulated, and for each Clifford instance the expectation values of four stabilizer and four destabilizer operators are measured. The displayed curves show the worst-case values relevant for the benchmark, namely the minimum stabilizer expectation value and the maximum absolute destabilizer expectation value over all sampled operators and Clifford unitaries. Despite the limited number of measurements per circuit, the figure clearly illustrates the characteristic transition from successful to unsuccessful benchmark outcomes as the system size increases, highlighting the practical applicability of the protocol. This numerical experiment demonstrates how the Clifford Volume is determined in practice by identifying the largest system size for which both acceptance conditions are simultaneously satisfied. In the example shown, this transition occurs at approximately $n = 33$ qubits.

\begin{figure}[t]
    \centering
    \includegraphics[width=0.9\linewidth]{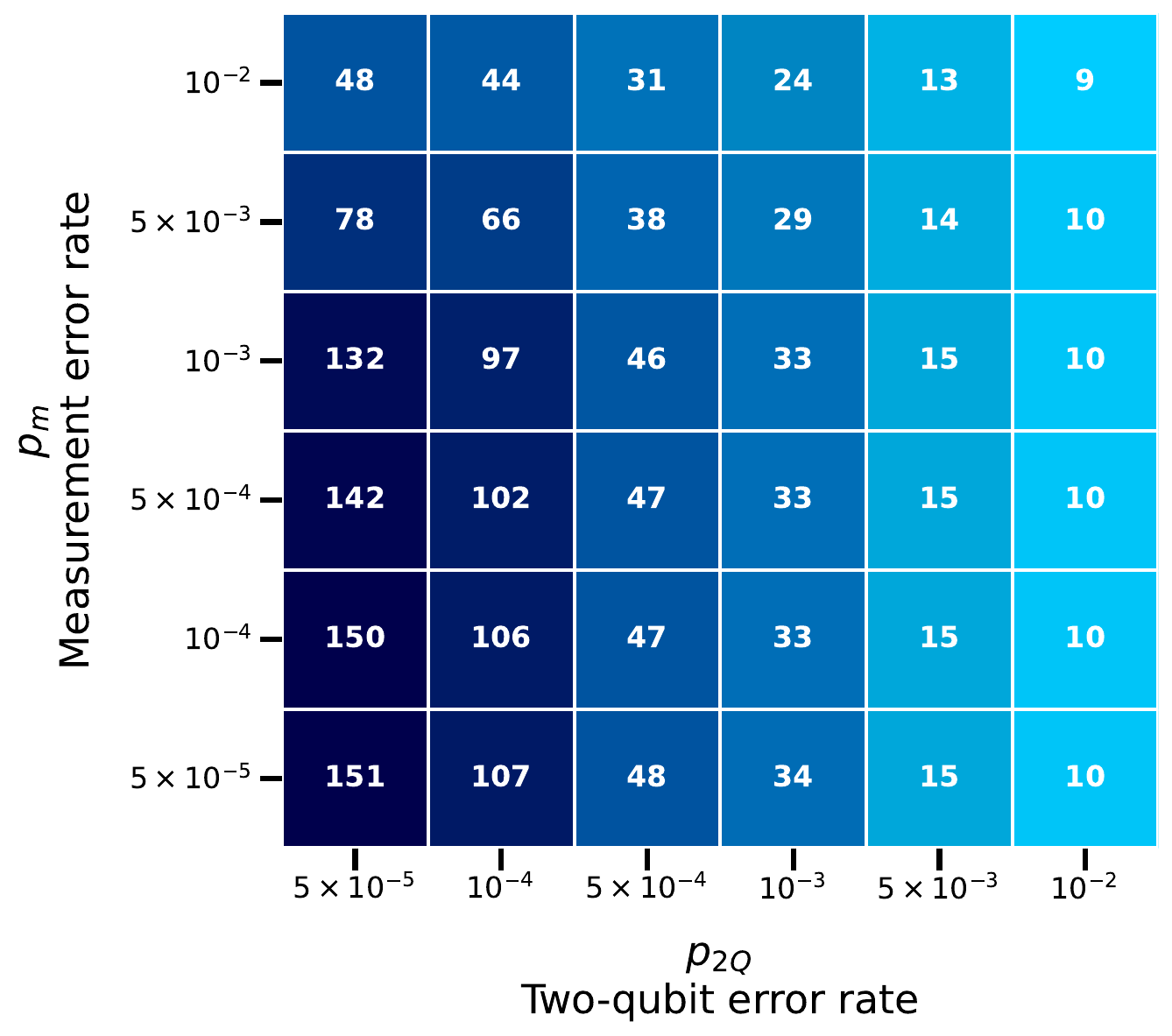}
    \caption{\label{fig:Clifford} Simulated CLV benchmark scores for different  pairs of error parameters. The horizontal axis corresponds to the two-qubit gate error probability $p_{2Q}$, and the vertical axis to the readout error probability $p_m$. Each cell indicates the largest qubit count $n$ for which the benchmark criteria are satisfied under the corresponding noise configuration.}
\end{figure}

Figure \ref{fig:Clifford} shows the CLV benchmark scores obtained by our simulations for different pairs of error parameters. It can be clearly seen that the benchmark yields reasonable values across a wide range of error parameters and is comparably sensitive to both readout and two-qubit gate errors.

\subsection{Realization of the Free Fermion Volume}

\begin{figure}[t]
    \centering
    \includegraphics[width=\linewidth]{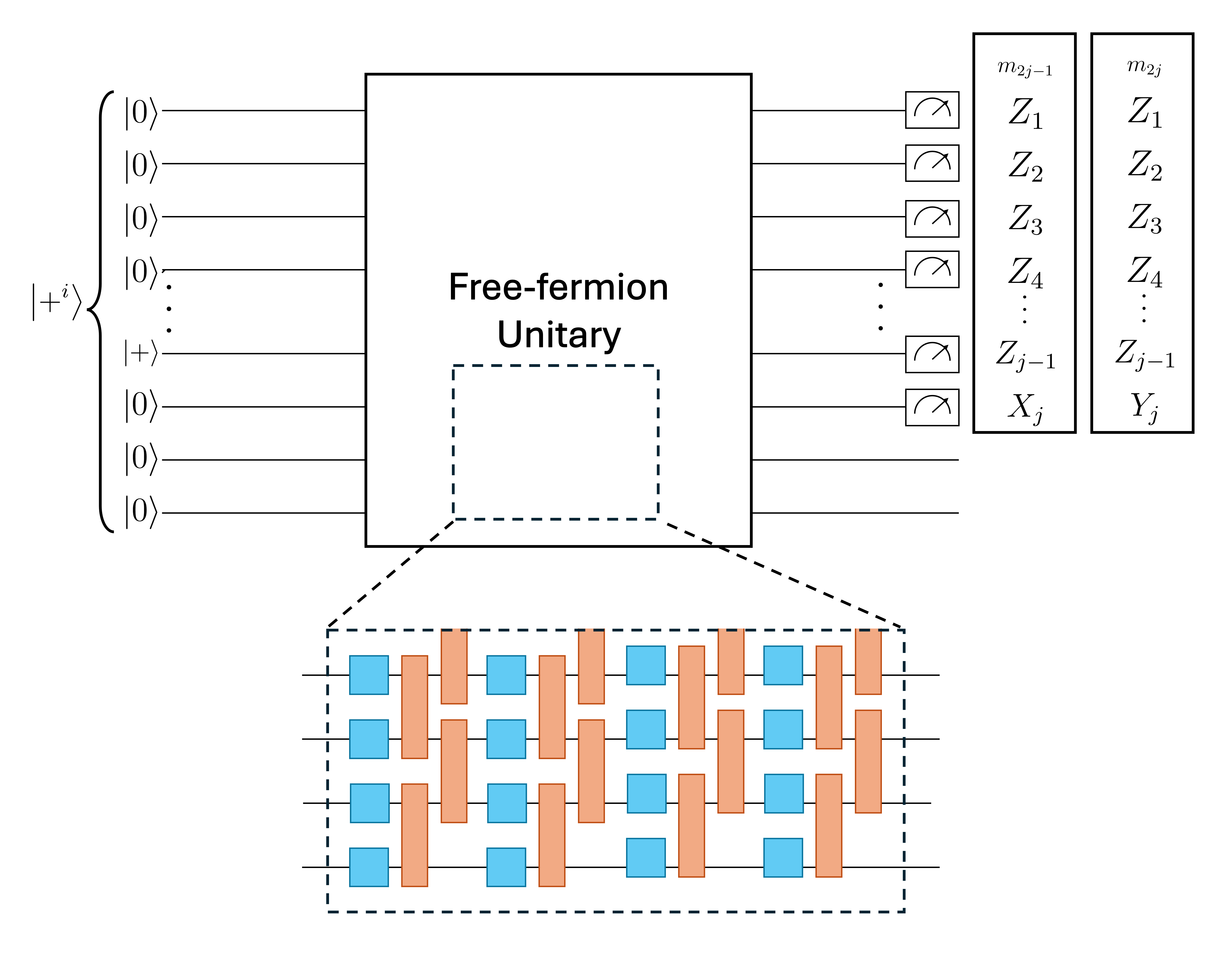}
    \caption{\label{fig : FreeFermion-Volume-Benchmark-scheme} Schematic representation of the Free Fermion Volume benchmark protocol. The circuit begins with all qubits initialized in the $|0\rangle$ state, except for a randomly chosen qubit prepared in the $|+\rangle$ state. This preparation distinguishes the corresponding Majorana operator, which -- unlike all others -- has an expectation value of $1$ in this state. A randomly sampled free-fermion unitary is then applied, decomposed into a brick-wall pattern of single- and two-qubit gates derived from Givens rotations of the corresponding $SO(2n)$ orthogonal matrix. After the unitary, single-qubit measurements are performed to  evaluate the expectation values of selected Majorana mode operators (mapped to Pauli operators on qubits) to assess the precision of the implementation and verify the orthogonality conditions of the underlying $SO(2n)$ transformation (see Appendix~\ref{FFV_step_by_step}).}
\end{figure}

In this section, we present a possible implementation of the FFV benchmark on the same virtual backend that was used in the CLV case. For this purpose, we developed a custom simulator specifically designed to efficiently handle the benchmarking process based on free-fermion operations. A schematic overview of the FFV benchmark protocol is shown in Fig.~\ref{fig : FreeFermion-Volume-Benchmark-scheme}. We assume that the virtual backend can directly implement the gate set $\{ H, S, \text{CNOT} \}$, complemented by parameterized single-qubit rotations $R_z(\alpha)$, as well as single-qubit measurements in the computational $Z$-basis. 

The first step in implementing the benchmark as a quantum circuit is to decompose a general free-fermion operation into elementary fermionic operations, which can then be mapped to qubit operations. To this end, we adopt a method for decomposing an arbitrary free-fermion operation into single- and two-qubit gates arranged in a brick-wall-like layout, as studied in Refs. \cite{Gluza2018, Oszmaniec2022}. Since the full $2^N$-dimensional unitary time evolution operator of a free fermion system can be compactly encoded in a $2N$-dimensional orthogonal matrix $O$, this allows us to construct the benchmark circuit by decomposing $O$, rather than the full unitary. The key insight enabling this approach is the standard decomposition of elements of the special orthogonal group $SO(2d)$ into a sequence of Givens rotations, which can then be translated into operations on qubits via an appropriate mapping.

The real (nearest-neighbor) Givens rotations have the form:
\begin{equation}
G^{(k)}(\alpha) =
\begin{bmatrix}
1 & \cdots & 0 & 0 & \cdots & 0 \\
\vdots & \ddots & \vdots & \vdots & \ddots & \vdots \\
0 & \cdots & \cos(\alpha) & -\sin(\alpha) & \cdots & 0 \\
0 & \cdots & \sin(\alpha) & \cos(\alpha) & \cdots & 0 \\
\vdots & \ddots & \vdots & \vdots & \ddots & \vdots \\
0 & \cdots & 0 & 0 & \cdots & 1
\end{bmatrix},
\end{equation}
where only the $2 \times 2$ block with row and column indices $k$ and $k+1$ is non-trivial. A general element $O \in SO(2d)$ can be decomposed in multiple ways. Here, we consider a decomposition where one alternates a series of nearest-neighbor Givens rotations $G^{(k)}$ with odd and even $k$ indices over $2d-1$ layers. In the $l$th layer, only real Givens rotations up to index $(d - |l-d|)$ are applied, and finally a diagonal matrix
\begin{equation}
S = \mathrm{diag}(s_1, s_2, \dots, s_d)
\end{equation}
with $s_i \in \{-1,1\}$ and $\prod_{i=1}^{d} s_i = 1$ is applied, ensuring that the method is naturally compatible with the nearest-neighbor topology. Thus, any special orthogonal matrix $O$ can be decomposed as
\begin{equation}
O = S \cdot \prod_{(k_i, \alpha_i) \in D} \left( G^{(k_i)}(\alpha_i) \right),
\label{eq : decomp}
\end{equation}
where $D$ defines the sequence of indices and rotation angles for the two-mode operations. Based on the definition of the free fermionic unitary operator, these rotations can then be mapped to qubit operations via the Jordan--Wigner transformation. For more details on the structure of the decomposition, see the appendix of Ref.~\cite{Oszmaniec2022}. This decomposition strategy produces a quantum circuit for the $n$-qubit free-fermion operation with $n(2n - 1)$ gates in total. Out of these $n(2n - 1)$ gates, $n(2n - 2)$ gates realize the Givens rotations, while the remaining $n$ gates form the final single-qubit layer. This results in a circuit with a typical two-qubit gate depth of approximately $2n - 1$.

We can handle the group of free-fermionic operators as the projective representation of the $SO(2d)$ group:
\begin{align}
\Pi: \quad &\text{SO}(2d) \to \text{U}\left[\mathcal{H}_{\text{Fock}}(\mathbb{C}^d)\right] \nonumber\\[1ex]
           &O \mapsto \exp\!\left( \frac{1}{4} \sum_{i,j=1}^{2d} \left[\log(O)\right]_{ij}\, m_i m_j \right).
\label{eq : map}
\end{align}
This representation provides a straightforward way to derive the decomposition of the free-fermion unitary. In particular, since $\Pi$ is a representation, it acts on the decomposed special orthogonal matrix $O$ as
\begin{equation}
\Pi(O) = \Pi(S) \; \prod_{(k_i, \alpha_i) \in D} \Pi\left( G^{(k_i)}(\alpha_i) \right).
\label{Pi(O)}
\end{equation}
A general Givens rotation is mapped to
\begin{align}
\Pi\!\left( G^{(k)}(\alpha) \!\right) & \!= \exp\!\left( \!\frac{1}{4} \sum_{i,j=1}^{2d} \!\left[\log\!\left( G^{(k)}(\alpha)\! \right)\!\right]_{ij} m_i m_j \!\right) \\[2mm] \notag
& = \exp\!\left( -\frac{\alpha}{2}\, m_k m_{k+1} \right),
\end{align}
where we used the anti-commutation relations of the Majorana operators, and the structure of the logarithm of the Givens rotation
\begin{equation}
\left[\log\!\left(\! G^{(k)}\!(\alpha) \!\right)\!\right]_{ij} \!\!=\!
\begin{cases}
-\alpha, \!&\!\!\!\! \text{if } (i,j)\! =\! (k\!+\!1, k), \\[1ex]
\alpha,  \!&\!\!\!\! \text{if } (i,j)\! =\! (k, k\!+\!1), \\[1ex]
0,       \!&\!\!\!\! \text{otherwise}.
\end{cases}
\end{equation}

We can determine the corresponding qubit transformation using the definition of the Majorana mode operators and the Jordan-Wigner transformation to map the fermionic operators to qubit operators. This transformation enables the simulation of fermionic systems using qubits in quantum circuits, as the unitary defined in Eq.~(\ref{eq : map}) becomes
\begin{equation}
U(2l-1,\alpha)
=
I^{\otimes(l-1)}
\otimes e^{-i\alpha Z_l/2}
\otimes I^{\otimes(N-l)} ,
\end{equation}

for odd indices $k=2l-1$, while

\begin{equation}
U(2l,\alpha)
=
I^{\otimes(l-1)}
\otimes e^{-i\alpha X_lX_{l+1}/2}
\otimes I^{\otimes(N-l-1)} ,
\end{equation}
for even indices $k=2l$. These expressions define two different types of unitaries: One- and two-qubit operations, depending on the parity of the chosen Majorana modes. Both types of operations can be implemented using simple quantum logic gates as follows: For odd-indexed Majorana modes ($k = 2l - 1$), the unitary reduces to a single-qubit rotation around the $z$-axis on the $l$th qubit $U(k, \alpha) = R_z(\alpha)$. For even-indexed Majorana modes ($k = 2l$), the unitary becomes a two-qubit operation that can be decomposed as $U(k, \alpha) = (H_l \otimes H_{l+1}) \cdot \text{CX}_{l,l+1} \cdot (I \otimes R_z(\alpha)) \cdot \text{CX}_{l,l+1} \cdot (H_l \otimes H_{l+1})$. This sequence combines single-qubit Hadamard gates, a rotation around the $z$-axis, and CNOT (CX) gates.

The remaining term in the decomposition in Eq.~(\ref{Pi(O)}) is the representation of the diagonal special orthogonal matrix $S$, which  can be realized as a single layer of single-qubit Pauli operators. It is straightforward to see the gate decomposition, starting from the relationships between the Majorana operators under the time evolution 
\begin{equation}
U m_i U^\dagger = \sum_{j=1}^{2d} S_{ij} m_j = s_i m_i \, ,
\end{equation}
which conveys that the unitary $U$ must commute with Majorana mode operators corresponding to a positive $s_i$ value, while for negative $s_i$ values, the $U$ must anti-commute. Since the Majorana mode operators anti-commute, the operator $U = \prod_{\{k : s_k = -1 \}} m_k$ meets the expected commutation and anti-commutation properties. Considering the Jordan-Wigner transformation, one can see that this unitary can be implemented as a layer of Pauli operators in the following way:
\begin{equation}
U = \prod_{\substack{k : s_k = -1 \\ \text{odd } k}} Z_1 \cdots Z_{p-1} X_k \cdot \prod_{\substack{k : s_k = -1 \\ \text{even } k}} Z_1 \cdots Z_{p-1} Y_k \, .
\end{equation}

The final step of the protocol is to measure the expectation values of all Majorana operators $ m_i $. The Majorana operators are mapped to qubit operations, and constructed using Pauli operators, thus their measurement involves appropriate transformations before applying computational basis measurements. Specifically, an $ H $ gate is applied to qubit $ i $ for $ X_i $, an $ S^\dagger $ gate followed by an $ H $ gate is applied to qubit $ i $ for $ Y_i $, and no additional gates are required for $ Z $-terms before measurements in the computational basis. From the obtained bitstrings we can compute the parity of the measurement outcomes for the corresponding Pauli string to determine the eigenvalue of $ m_i $. Then, in the usual way, we average over multiple measurements to obtain the expectation value.

\begin{figure}[h]
\centering
\includegraphics[width=\linewidth]{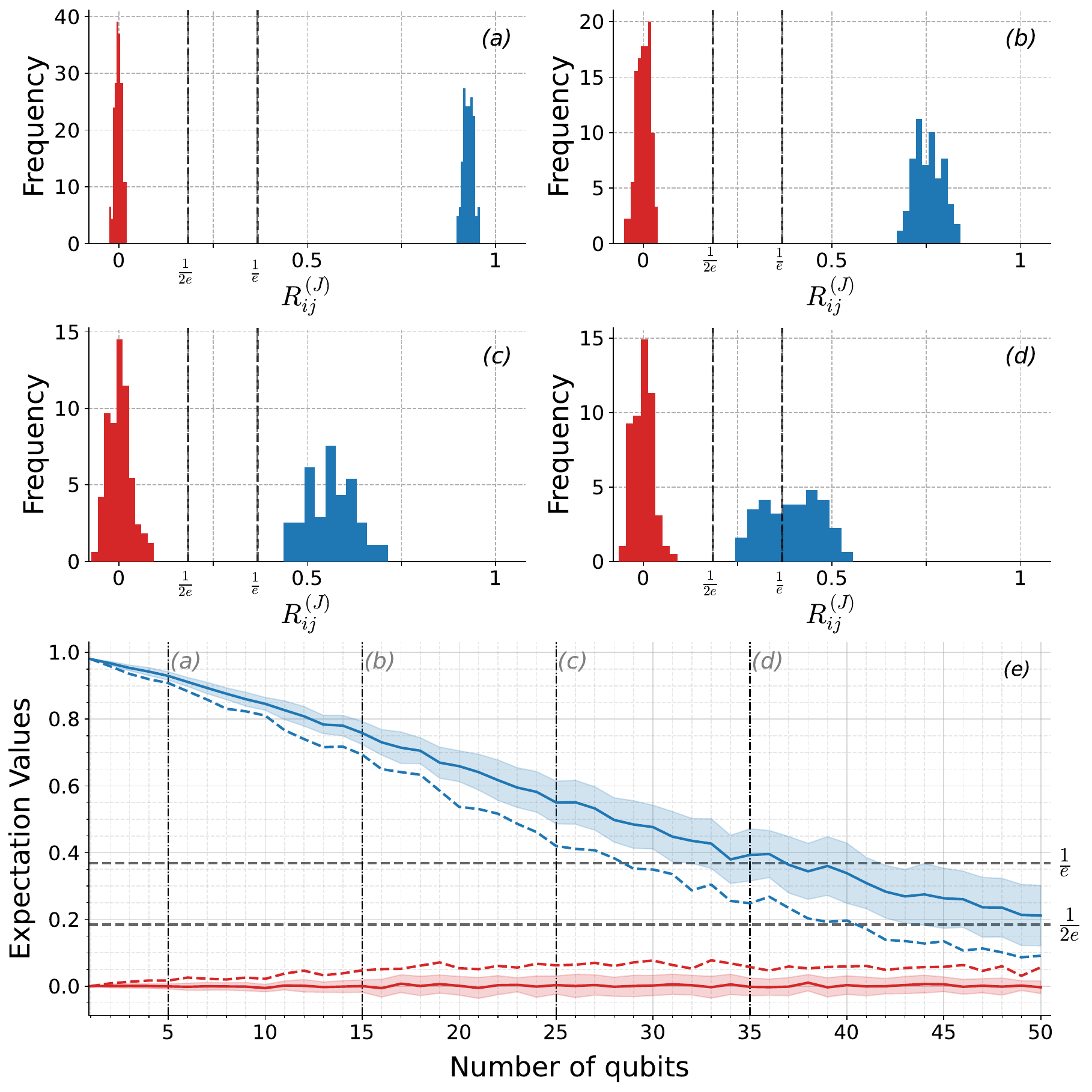}
\caption{
Distribution of simulated measured expectation values of Majorana mode operators mapped to Pauli operators acting on qubits in the FFV benchmark, with a readout error of $p_m = 10^{-2}$ and two-qubit gate error of $p_{2Q} = 10^{-3}$. Subfigures (a)–(d) show histograms of the values $R_{ij}^{(J)}$ (defined in Eq.~\ref{reduced_lincomb}) for the parallel ($i = j$, blue) and orthogonal ($i \neq j$, red) linear combinations at qubit counts $n = 5$, 15, 25, and 35, respectively. Dashed lines indicate the thresholds: $\frac{1}{e}$ for the parallel condition and $\frac{1}{2e}$ for the orthogonal condition. Subfigure (e) shows the average values of the parallel (blue) and orthogonal (red) combinations as a function of qubit number. Shaded regions represent the standard deviation, dashed blue and red curves indicate the maximum deviation from the ideal values within the ensemble. Vertical dashed lines highlight qubit counts corresponding to subfigures (a)–(d).
}
\label{fig : Fermion}
\end{figure}

In order to demonstrate what FFV benchmark scores would be attainable in a realistic scenario, we numerically simulate the benchmark circuits with a well-controlled error model. We implement the same simple error model that was used for the CLV simulations; that is, after each two-qubit gate a randomly chosen two-qubit Pauli operator is applied with probability $ p_{2Q} $ which corresponds exactly to the widely considered two-qubit depolarization channel. For readout errors, we use a simple independent bit-flip error with probability $ p_{m}$. The behavior of the reduced orthogonality estimators $R_{ij}^{(J)}$ under this noise model is illustrated in Fig.~\ref{fig : Fermion}.

According to Eq.~(\ref{eq : expected}) two-qubit errors can be incorporated by determining the matrix elements of the corresponding time evolution operator $O$. After each term in the decomposition of the operator corresponding to a two-qubit gate, i.e., Majorana operators coupling to different sites, we apply an $SO(2n)$ operator describing the effect of the unintended Pauli operator. We can formulate this process with a quantum channel $\mathcal{E}^*$ acting on the Majorana operator, where, with probability $p_{2Q}$, the additional operators (manifesting the noise) are applied
\begin{align}
\mathcal{E}^*\left(m_q(0)\right) = &(1-p_{2Q}) \, U^\dagger m_q(0) U \\[3mm] \notag
& + \frac{p_{2Q}}{15}\sum_{P \in \mathcal{P}_2\setminus\{I\otimes I\}} P^\dagger U^\dagger m_q(0)  U P, 
\end{align}
where $\mathcal{P}_2$ denotes the set of two-qubit Pauli operators. Considering the free-fermionic nature of the time evolution, we can translate this channel to the language of special orthogonal operators acting on the Majorana operators
\begin{align}
m_q(t) &=  \mathcal{E}^*(m_q(0)) \notag \\ 
       &=  (1-p_{2Q}) \sum_k O_{q,k}(t)\, m_k(0) \\  
       & + p_{2Q} \sum_k \left[O_{\text{noisy}}(t)\right]_{q,k}\, m_k(0) \notag \,, 
\end{align}
which can be rewritten as 
\begin{equation}
\mathcal{E}^*(m_q(0)) = \sum_k \left[ O_{\text{eff}}\right]_{q,k} m_k(0) \, ,
\end{equation}
where $O_{\text{eff}} = (1-p_{2Q})O + p_{2Q}\,O_{\text{noisy}}$, the probability-weighted sum of the error-free and noisy time-evolution operators that captures the effect of noise on the time evolution. In order to define the effective operator $O_{\text{eff}}$, we need to determine the $SO(2n)$ operator $O_{\text{noisy}}$ that encodes the effect of the Pauli errors. Fortunately, this is straightforward to do based on the commutation properties of the Majorana operators. For example, the overall effect of an $X$ error on the $i$th qubit, corresponding to the two-qubit Pauli operator $X_i I_{i+1}$ in the computational basis, can be encoded as a diagonal transformation acting on the Majorana operators:
\begin{equation}
O_{XI} = \mathrm{diag}\left(\underbrace{1,\ldots,1}_{2i-1},\,-1,\underbrace{-1,\ldots,-1}_{2n-2i}\right).
\end{equation}
This operator introduces a sign flip for the Majorana operators at sites where the error does not commute with the corresponding qubit operator. By determining for each possible error $P$ the corresponding $O_{P}$ transformation, one can write the $O_{\text{noisy}}$ operator as
\begin{equation}
O_{\text{noisy}} = \frac{1}{15}\sum_{P \in \mathcal{P}_2\setminus\{I\otimes I\}} O_{P} \; O \,.
\end{equation}
Based on the effective time evolution operator, it is straightforward to determine the expectation values in the presence of noise
\begin{equation}
\braket{m_j'}_{\rho_i} = \left[O_{\text{eff}}\right]_{j,i} \, .    
\end{equation}

In order to take into account the effect of readout errors, we incorporate the same simple error model that we used for the simulation of the CLV benchmark. We assume that in the experiment only $Z$-basis measurements are available. Therefore, for operators that are not diagonal in the $Z$-basis, a basis transformation is applied on the qubit on which an $ X $ or $Y$ operator appears. This maps the operator on that qubit to $Z$ so that both $m_{2i-1}$ and $m_{2i}$ are transformed to the effective measurement operator
\begin{equation}
\tilde{m}_q = Z_1 Z_2 \cdots Z_i,
\end{equation}
which acts nontrivially on the first $i$ qubits. In order to simulate measurement errors, an $X$ operator is applied independently on each qubit with probability $p_{\text{m}}$, so that each qubit undergoes the error
\begin{equation}
E_j = 
\begin{cases} 
I & \text{with probability } 1-p_{\text{m}} \\
X & \text{with probability } p_{\text{m}} \, .
\end{cases}
\end{equation}
Since the measurement is performed in the $Z$ basis after the basis change, an $X$ error on a qubit that is measured as $Z$ will flip its sign, contributing with a multiplicative factor of $(1-2p_{m})$. Thus, for an operator acting non-trivially on $i$ qubits, the measured expectation value in the presence of errors becomes
\begin{equation}
\braket{ m_j' }_{\rho_i,\text{error}} = (1-2p_\text{m})^{\lceil \frac{j}{2} \rceil}\, \braket{ m_j' }_{\rho_i} \, ,
\end{equation}
which shows that the simulated bit-flip errors cause an effective damping of the expectation values by a factor of $(1-2p_{m})$ for each qubit on which the measurement operator acts non-trivially.

In order to evaluate the performance of the FFV benchmark, we simulated its instances for realistic error parameters. For each parameter pair, we computed the expectation values of selected Majorana mode operators mapped to Pauli operators, using ensembles of random free-fermion circuits. As shown in Fig. \ref{fig : Fermion}, the expectation values for the "parallel" linear combinations ($i = j$) decay with increasing qubit count, gradually approaching the predefined threshold. Meanwhile, the "orthogonal" combinations ($i \ne j$) remain centered near zero, but exhibit growing standard deviations and increasingly large outliers. The second benchmark criteria of Eq.~(\ref{FFV_criteria}) for extreme deviations have been set out to capture these effects and make the FFV benchmark sensitive to noise in both cases.

\begin{figure}[h]
    \centering
    \includegraphics[width=0.9\linewidth]{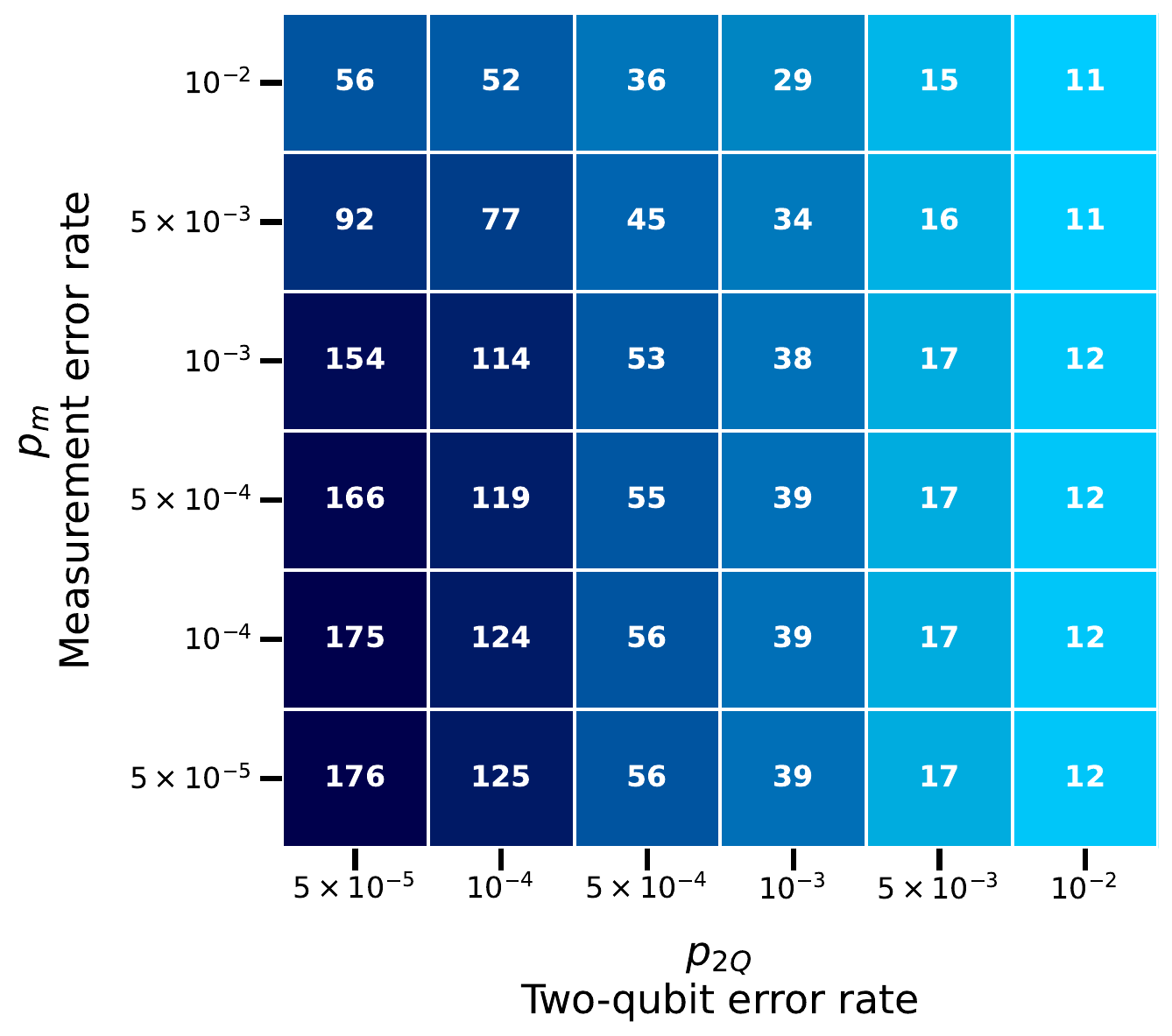}
    \caption{FFV benchmark scores for different pairs of error parameters. The horizontal axis corresponds to the two-qubit gate error probability $p_{2Q}$, and the vertical axis to the readout error probability $p_m$. Each cell indicates the largest qubit count $n$ for which the FFV benchmark criteria are satisfied.
    }
    \label{fig : Fermion-Volume}
\end{figure}

The heatmap of Fig.~\ref{fig : Fermion-Volume} shows the maximum qubit count for which the benchmark criteria are satisfied under each noise configuration, thereby corresponding to a hypothetical FFV benchmark score that would be achievable with the simulated virtual device. 

The results demonstrate that the FFV benchmark behaves consistently and is comparably sensitive to both readout ($p_m$) and two-qubit gate errors ($p_{2Q}$). In general, lower error rates allow the benchmark to scale to larger qubit counts. The dependence is approximately symmetric with respect to the two error sources, indicating that the benchmark is similarly constrained by both. For the lowest error rates tested ($p_{2Q} \leq 10^{-4}$, $p_m \leq 10^{-3}$), the benchmark supports system sizes exceeding $100$ qubits, while in high-noise settings ($p_{2Q}, p_m \geq 10^{-2}$), the score is close to $10$. This smooth and interpretable scaling behavior confirms that the FFV benchmark can reliably capture the interplay between computational capability and hardware noise, confirming its utility for characterizing quantum hardware in the NISQ and late-NISQ regime.

\FloatBarrier
\section{Demonstration of the CLV benchmark on Quantinuum Hardware}
\label{sec: Demonstration}

In what follows, we present experimental data demonstrating the applicability and feasibility of the Clifford Volume benchmark on real quantum hardware. The experiments were carried out on the H2-1 model of a trapped-ion quantum computer of Quantinuum. The benchmark circuits used in this demonstration were generated and evaluated using the open-source software introduced in Ref.~\cite{EQCB}, which is publicly available. A detailed characterization of the device is beyond the scope of this work; however, we note that the platform is specified by nominal single- and two-qubit gate error rates and measurement errors representative of state-of-the-art trapped-ion systems. Based on these nominal error rates and our simple numerical simulations (see Fig. \ref{fig:Clifford}), one expects a CLV on the order of a few tens. In particular, the H2-1 specifications suggest an achievable CLV of approximately thirty, which served as a starting point for the experimental search region.

Let us point out that the aim of these experiments was not to provide the most optimized or hardware-specific performance assessment. Rather, the goal was to experimentally validate the proposed benchmark methodology and to establish a baseline CLV value for this platform, which may be improved further through dedicated hardware-aware compilation techniques and parameter tuning. Throughout the experiments, circuit compilation was performed using the tket-based compiler provided and hosted by Quantinuum, configured for the H2-1 platform, with the highest available compiler optimization level selected. Further details on the compilation of the benchmark circuits are provided in Appendix~\ref{appendix : circuit_compilation}.

\begin{figure}[h]
    \centering
    \includegraphics[width=\linewidth]{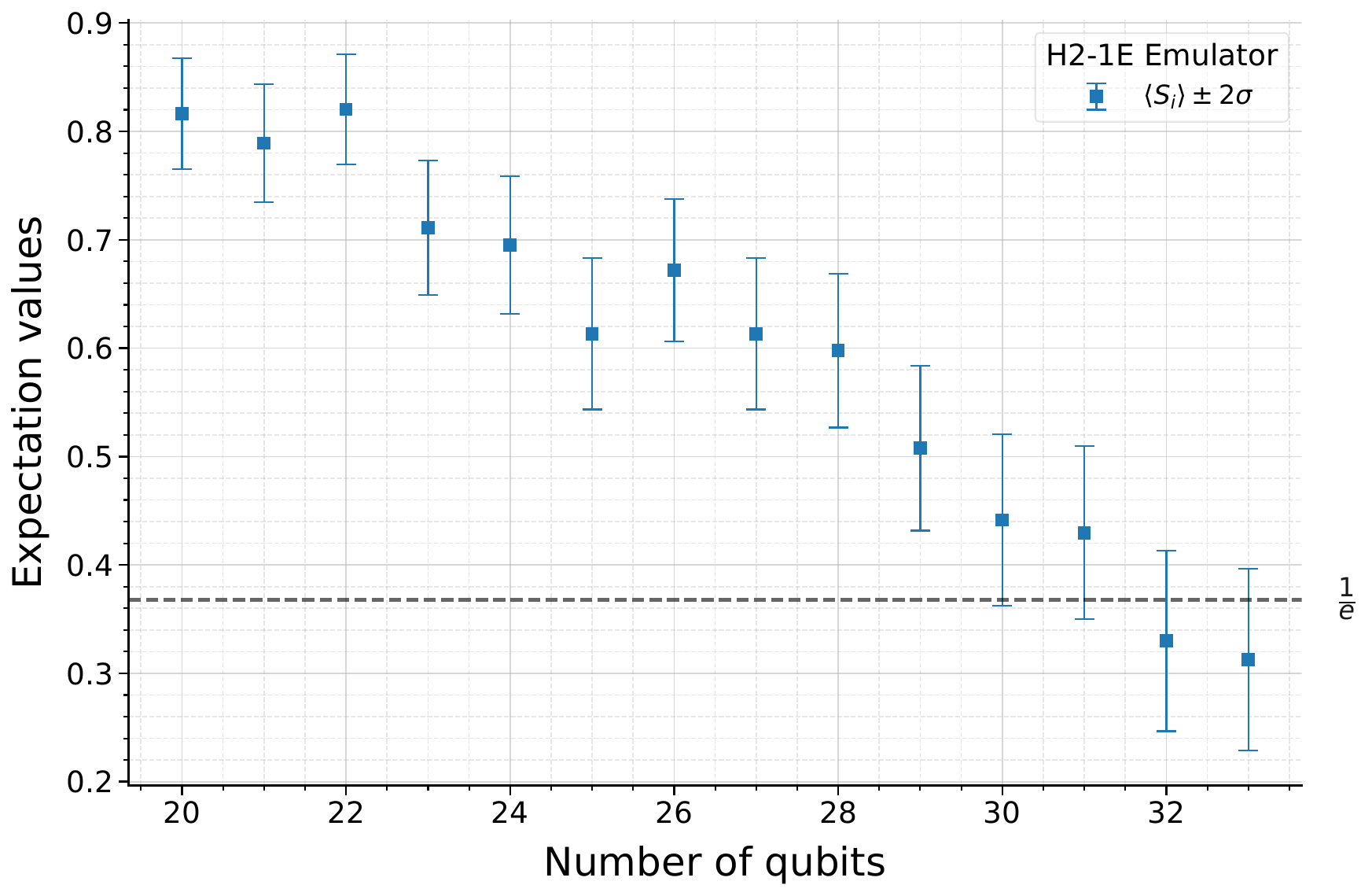}
    \caption{The lowest simulated stabilizer expectation values obtained on Quantinuum's H2-1 emulator backend as a function of the number of qubits. For each qubit count $(n)$, the plotted value is the minimum expectation value observed among all stabilizer measurements performed for the sampled $n$-qubit Clifford circuit. The number of stabilizer operators measured at each qubit count was adjusted based on the numerical simulation results, in order to efficiently probe the relevant performance regime. Error bars indicate the statistical uncertainty of the measured expectation value ($\pm 2\sigma_{\mathcal{S}})$ arising from finite sampling. The horizontal dashed line marks the stabilizer acceptance threshold of $\frac{1}{e}$.}
    \label{fig : emulator_scan}
\end{figure}

As the focus of this work was on validation of the benchmark process rather than hardware optimization, our experimental strategy was guided by results obtained from the hardware emulators provided by Quantinuum (see Fig.~\ref{fig : emulator_scan} presenting the lowest stabilizer expectation values obtained). 

\begin{figure}[h]
    \centering
    \includegraphics[width=\linewidth]{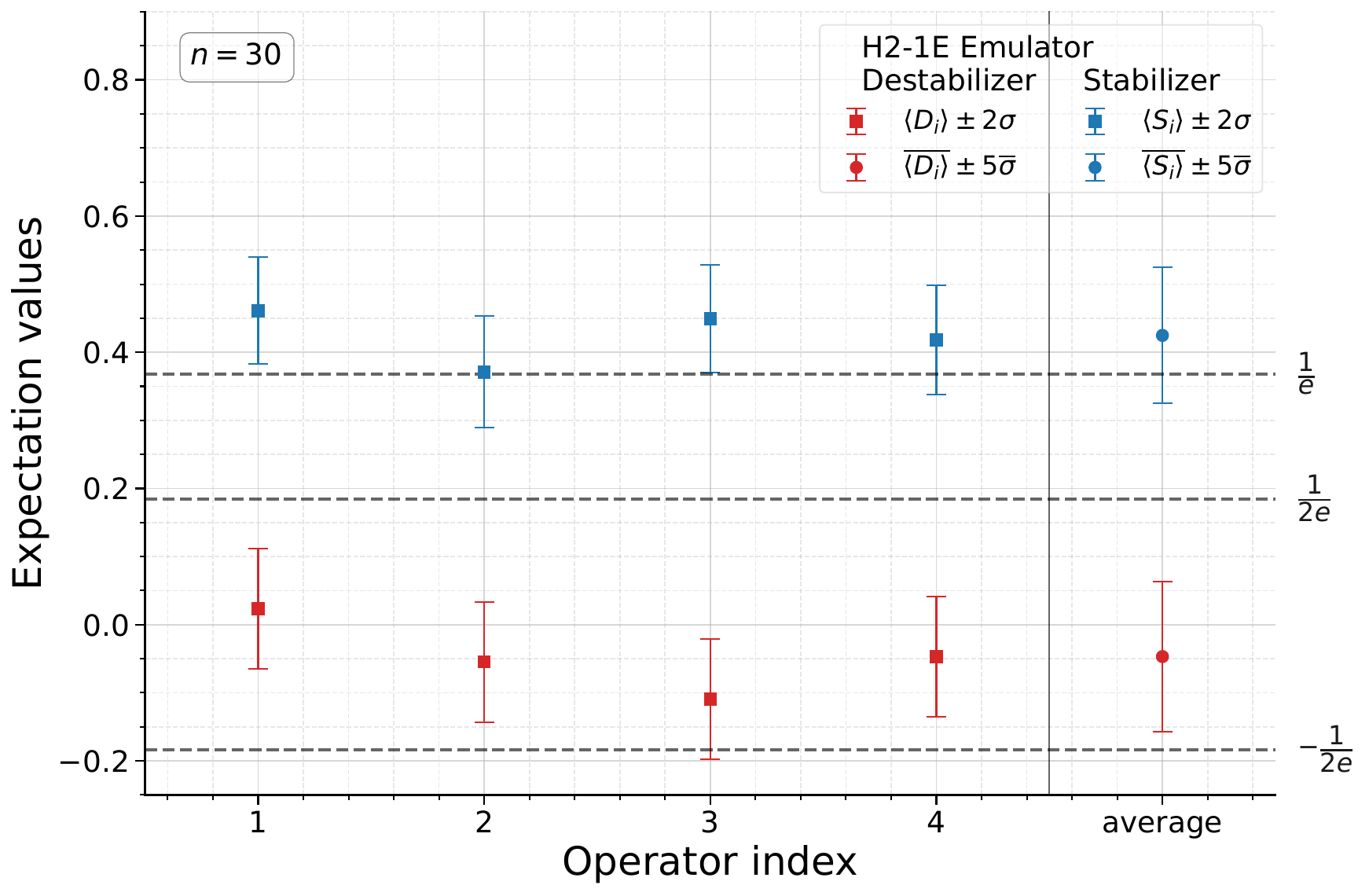}
    \caption{Evaluation of a single CLV benchmark measurement round on the emulator backend for $n = 30$ qubits corresponding to the first Clifford operator that fails the benchmark criterion. For this Clifford instance, expectation values of four stabilizer ($S_i$) and four destabilizer ($D_i$) operators are shown. Each expectation value was estimated from $512$ measurement shots. Points indicate the measured values, with error bars representing statistical uncertainties ($\pm 2\sigma$ and $\pm 5\sigma$) arising from finite sampling. The rightmost point shows the average stabilizer expectation value used to determine benchmark success or failure. Horizontal dashed lines indicate the acceptance thresholds $\frac{1}{e}$ and $\frac{1}{2e}$. The corresponding expectation values are shown in Table~\ref{tab : pauli_expectations_30}.}
    \label{fig : clifford_30}
\end{figure}

The emulators were used to identify the range of qubit counts for which the device is expected to reliably implement random Clifford unitaries according to the benchmark criteria (see also Fig.~\ref{fig : clifford_30}, which shows that at qubit count $n=30$, the emulated results suggest a possible failure). The analysis on the emulators enabled us to restrict the experimental search space. Since the success conditions are formulated in a worst-case manner, we stopped the evaluation whenever a measured expectation value violated the benchmark criteria at a given qubit count, and the benchmark was subsequently repeated for smaller qubit counts. We first performed a rough scan over the emulator-identified region using a reduced number of stabilizer operators for each candidate qubit count, which reduced the experimental resource requirements. This procedure enabled an educated estimate of the achievable Clifford Volume, which was then refined by more detailed measurements in the vicinity of the expected threshold. 

\begin{table}[h]
\centering
\small
\begin{tabular}{|c|c|c|}
\hline
\makecell{$i$}
& \makecell{Stabilizer \\ $\langle S_i\rangle \pm 2\sigma_{S_i}$}
& \makecell{Destabilizer \\ $\langle D_i\rangle \pm 2\sigma_{D_i}$} \\
\hline\hline
1 & $0.4609 \pm 0.078$ & $0.0234 \pm 0.088$ \\
2 & $0.3711 \pm 0.082$ & $-0.0547 \pm 0.088$ \\
3 & $0.4492 \pm 0.079$ & $-0.1094 \pm 0.088$ \\
4 & $0.4180 \pm 0.080$ & $-0.0469 \pm 0.088$ \\
\hline

\end{tabular}
\caption{Expectation values of stabilizer and destabilizer operators shown in Fig.~\ref{fig : clifford_30}. For the explicit form of the measured Pauli operators see Table~\ref{tab : pauli_strings_30}.}
\label{tab : pauli_expectations_30}
\end{table}

Using the emulator estimates, we initially predicted that the device would exhibit a CLV of approximately 30–31 (see Fig. ~\ref{fig : clifford_30}). The experimental results, however, indicated that the real device slightly outperforms the emulator predictions for this specific task. 

\begin{figure}[h]
    \centering
    \includegraphics[width=\linewidth]{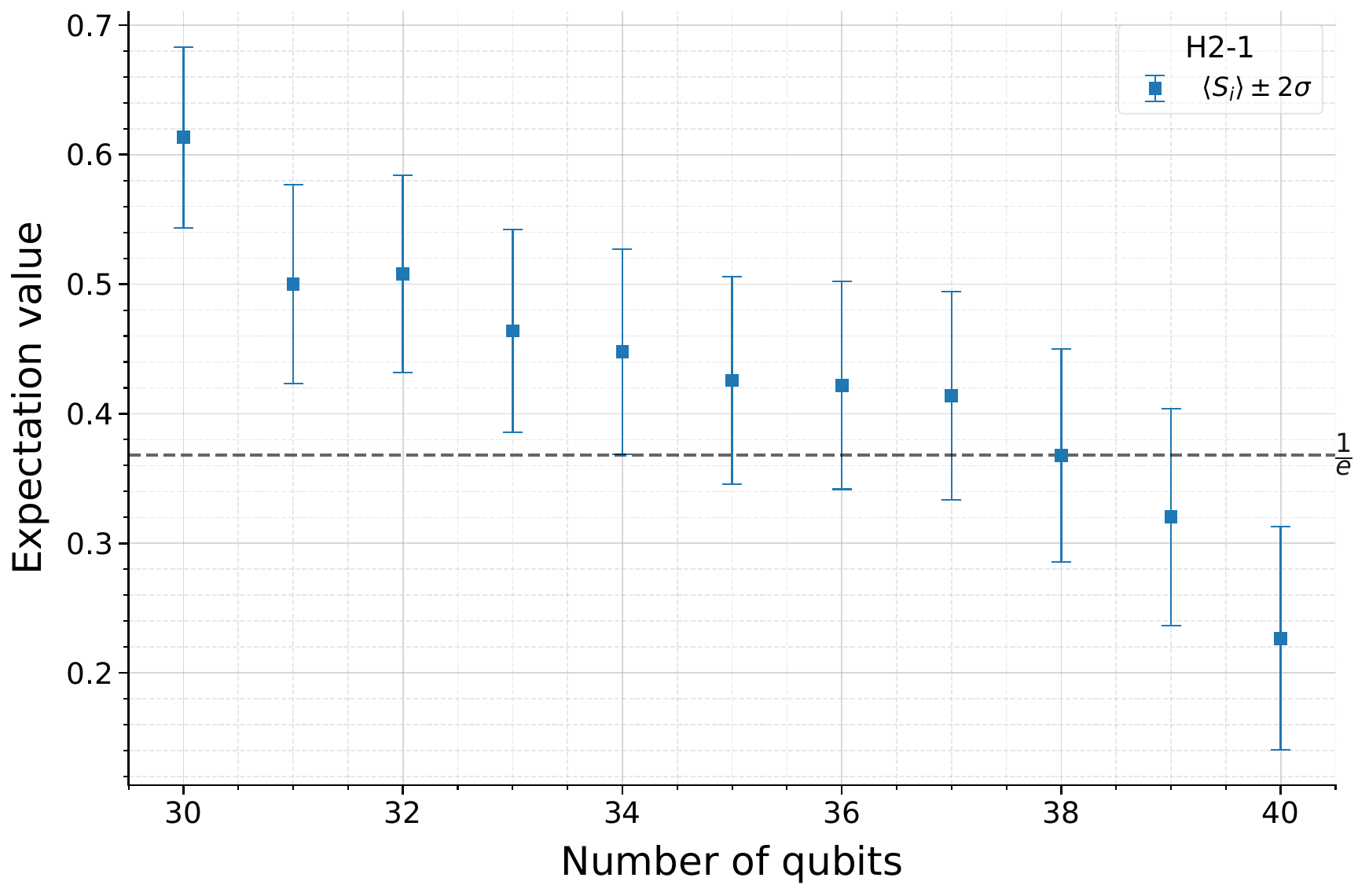}
    \caption{Results collected during the initial scan of the CLV benchmark on  Quantinuum H2-1. For each qubit count the expectation value of a single stabilizer operator associated with a randomly chosen Clifford unitary was measured. Each data point corresponds to $512$ measurement shots, with error bars indicating  statistical uncertainty of $2\sigma_{\mathcal{S}}$, where $\sigma_{\mathcal{S}}$ is given in Eq.~(\ref{sigma_S_D}). The dashed horizontal line marks the stabilizer acceptance threshold $\frac{1}{e}$.
    }
    \label{fig : clifford_scan}
\end{figure}

An initial scan was therefore performed by measuring a single stabilizer associated with a randomly chosen Clifford unitary for each qubit count in the interval from 30 to 40 qubits, which allowed us to localize the relevant region of interest (see Fig. ~\ref{fig : clifford_scan}). 

\begin{figure}[h]
    \centering

    \includegraphics[width=\linewidth]{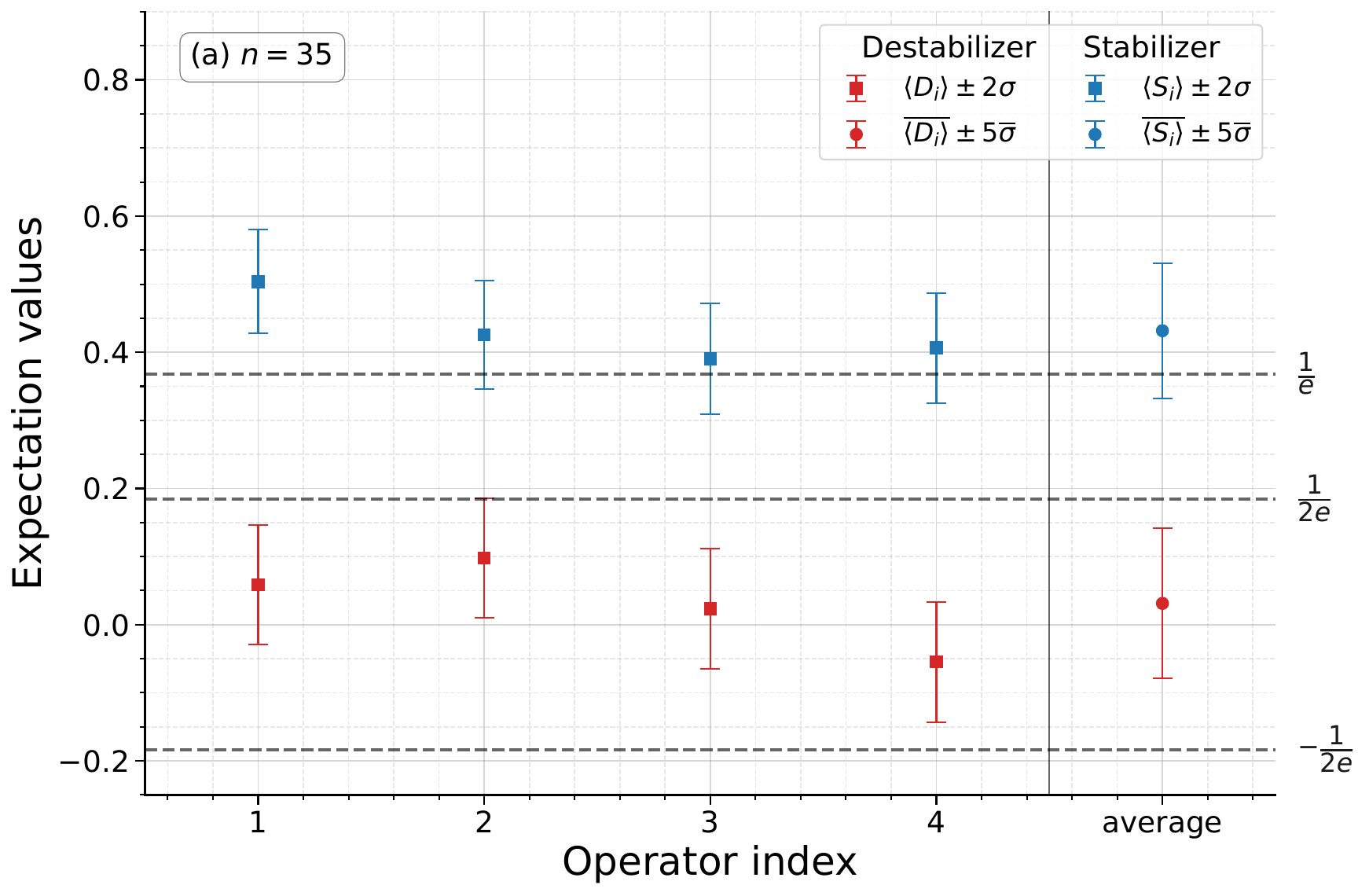}

    \vspace{0.5em}

    \includegraphics[width=\linewidth]{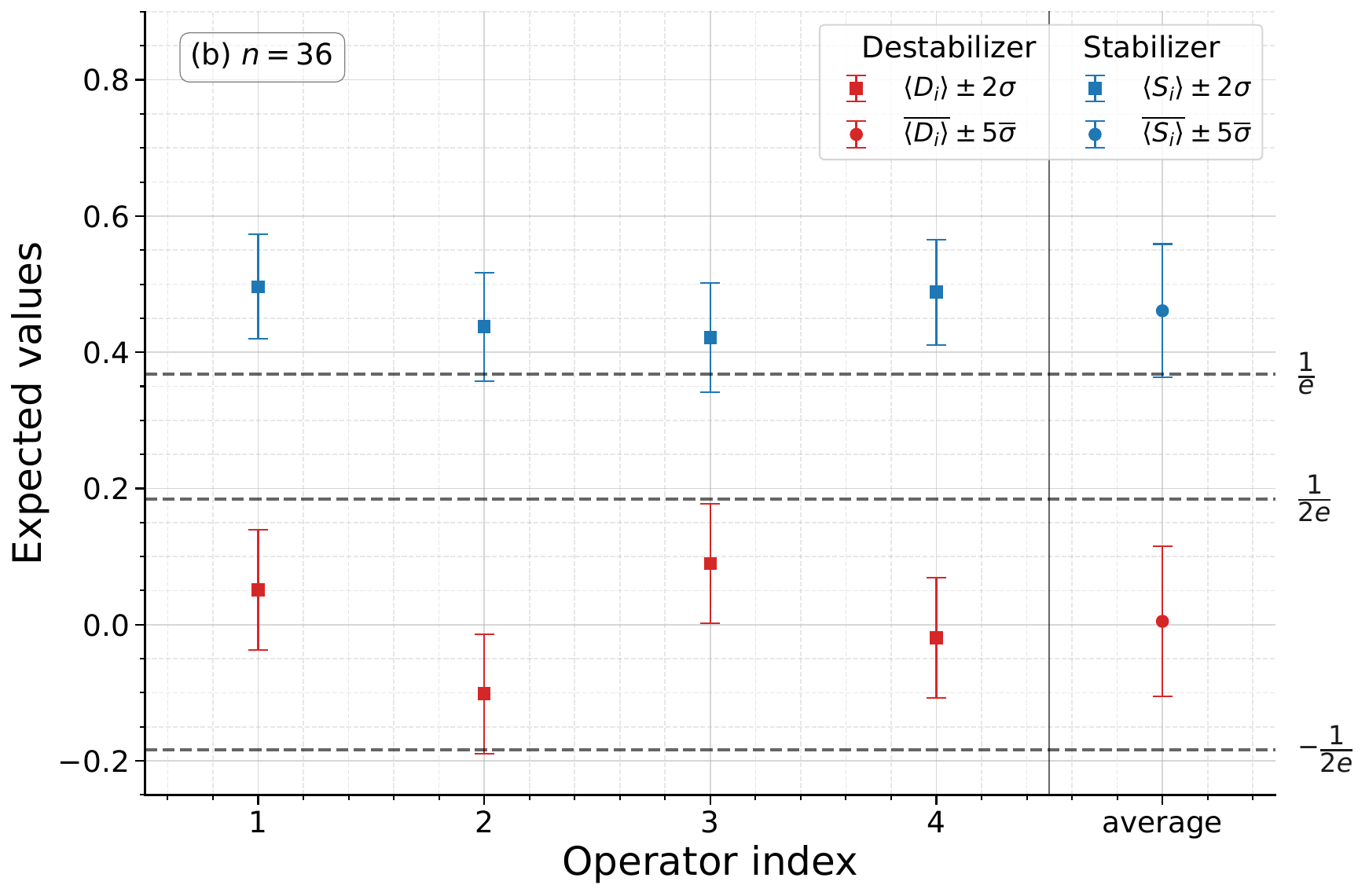}

    \caption{CLV benchmark results for a single measurement round on Quantinuum H2-1 at qubit counts (a) $n=35$ and (b) $n=36$. For each qubit count, the expectation values of four stabilizer operators (blue) and four destabilizer operators (red), associated with a single randomly chosen Clifford unitary, are shown. Error bars indicate statistical uncertainty of $2\sigma_{\mathcal{P}}$, where $\sigma_{\mathcal{P}}$ is given by Eq.~(\ref{sigma_S_D}). The rightmost marker represents the average of the expectation values, with error bars indicating $5\overline{\sigma}_{\mathcal{P}}$ uncertainty, where $\overline{\sigma}_{\mathcal{P}}$ is given by Eq.~(\ref{sigma_S_D_avg}). Each expectation value was estimated using 512 measurement shots. Dashed horizontal lines denote the benchmark acceptance thresholds. The corresponding expectation values are shown in Table~\ref{tab:pauli_expectations_35_36}.}
    \label{fig:clv_n35_n36}
\end{figure}

Based on these results, a full round of the CLV benchmark was carried out for qubit counts near the identified threshold. For qubit counts of 35 and 36, the defined success criteria were clearly violated (see Fig.~\ref{fig:clv_n35_n36} and Table~\ref{tab:pauli_expectations_35_36}).

\begin{figure}[t]
    \centering
    \includegraphics[width=\linewidth]{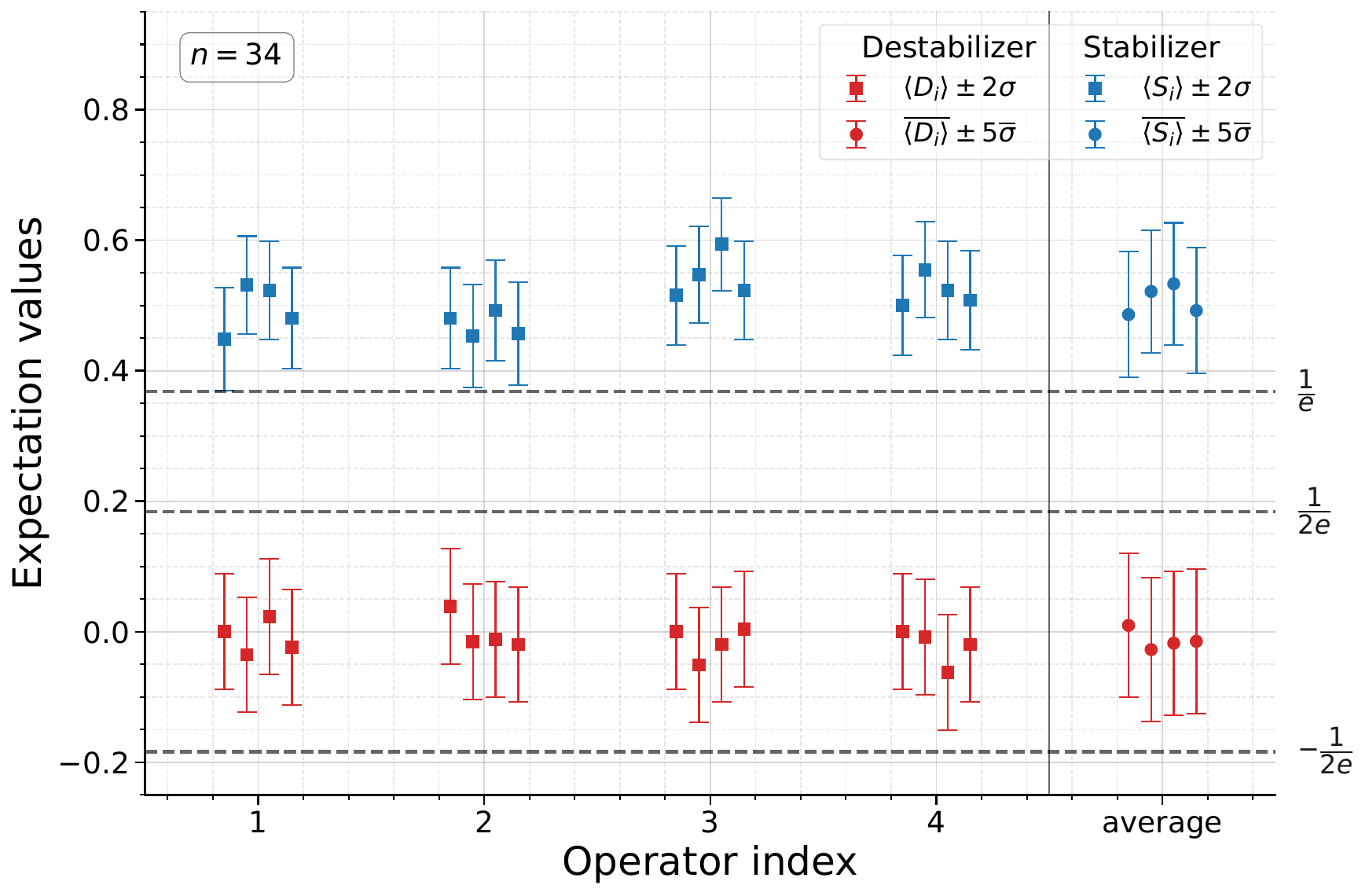}
    \caption{CLV benchmark results for a qubit count of $34$ on  Quantinuum H2-1. Measured expectation values of four stabilizer (blue) and four destabilizer operators (red) are shown for four randomly chosen Clifford unitaries, obtained over four independent measurement rounds. Markers corresponding to operators with the same index across different rounds are shown adjacent to each other. Each expectation value was estimated using 512 measurement shots. All four rounds satisfy the benchmark criteria, establishing $34$ as the CLV of the device under the conditions of this experiment. Corresponding expectation values are shown in Table~\ref{tab : Clifford_34}.}
    \label{fig : clifford_34}
\end{figure}

\begin{table}[h]
\centering
\small

\begin{tabular}{|c|c|c|}
\hline
\makecell{$i$}
& \makecell{Stabilizer \\ $\langle S_i\rangle \pm 2\sigma_{S_i}$}
& \makecell{Destabilizer \\ $\langle D_i\rangle \pm 2\sigma_{D_i}$} \\
\hline\hline
1 & $0.504 \pm 0.076$ & $0.059 \pm 0.088$ \\
2 & $0.426 \pm 0.080$ & $0.098 \pm 0.088$ \\
3 & $0.391 \pm 0.081$ & $0.023 \pm 0.088$ \\
4 & $0.406 \pm 0.081$ & $-0.055 \pm 0.088$ \\
\hline
\end{tabular}

\vspace{0.3em}
\textbf{(a) $n=35$}

\vspace{1.0em}

\begin{tabular}{|c|c|c|}
\hline
\makecell{$i$}
& \makecell{Stabilizer \\ $\langle S_i\rangle \pm 2\sigma_{S_i}$}
& \makecell{Destabilizer \\ $\langle D_i\rangle \pm 2\sigma_{D_i}$} \\
\hline\hline
1 & $0.496 \pm 0.077$ & $0.051 \pm 0.088$ \\
2 & $0.438 \pm 0.079$ & $-0.102 \pm 0.088$ \\
3 & $0.422 \pm 0.080$ & $0.090 \pm 0.088$ \\
4 & $0.488 \pm 0.077$ & $-0.020 \pm 0.088$ \\
\hline
\end{tabular}

\vspace{0.3em}
\textbf{(b) $n=36$}

\caption{Expectation values of stabilizer and destabilizer operators shown in Fig.~\ref{fig:clv_n35_n36} for qubit counts (a) $n=35$ and (b) $n=36$. For the explicit form of the measured Pauli operators, see Tables~\ref{tab : pauli_strings_35} and \ref{tab : pauli_strings_36}, respectively.
}
\label{tab:pauli_expectations_35_36}
\end{table}
In contrast, for 34 qubits the measured expectation values were found to lie very close to the threshold, but across all four measurement rounds the system consistently satisfied the benchmark criteria. We therefore assign a Clifford Volume of 34 to the H2-1 device under the conditions of this experiment (see Fig. ~\ref{fig : clifford_34} and  Table ~\ref{tab : Clifford_34}).

\begin{table}[t]
\centering
\small
\begin{tabular}{|c|c|c|c|}
\hline
\makecell{$\mathcal{C}^k$}
& \makecell{$i$}
& \makecell{Stabilizer \\ $\braket{\mathcal{S}^k_i} \pm 2\sigma_{\mathcal{S}^k_i}$}
& \makecell{Destabilizer \\ $\braket{\mathcal{D}^k_i} \pm 2\sigma_{\mathcal{D}^k_i}$} \\
\hline \hline

\multirow{4}{*}{$\mathcal{C}^1$}
& 1 & $0.448 \pm 0.079$ & $0.000 \pm 0.088$ \\
& 2 & $0.480 \pm 0.078$ & $0.039 \pm 0.088$ \\
& 3 & $0.516 \pm 0.076$ & $0.000 \pm 0.088$ \\
& 4 & $0.500 \pm 0.077$ & $0.000 \pm 0.088$ \\
\hline

\multirow{4}{*}{$\mathcal{C}^2$}
& 1 & $0.531 \pm 0.075$ & $-0.035 \pm 0.088$ \\
& 2 & $0.453 \pm 0.079$ & $-0.016 \pm 0.088$ \\
& 3 & $0.547 \pm 0.074$ & $-0.051 \pm 0.088$ \\
& 4 & $0.555 \pm 0.074$ & $-0.008 \pm 0.088$ \\
\hline

\multirow{4}{*}{$\mathcal{C}^3$}
& 1 & $0.523 \pm 0.075$ & $0.023 \pm 0.088$ \\
& 2 & $0.492 \pm 0.077$ & $-0.012 \pm 0.088$ \\
& 3 & $0.594 \pm 0.071$ & $-0.020 \pm 0.088$ \\
& 4 & $0.523 \pm 0.075$ & $-0.063 \pm 0.088$ \\
\hline

\multirow{4}{*}{$\mathcal{C}^4$}
& 1 & $0.480 \pm 0.078$ & $-0.023 \pm 0.088$ \\
& 2 & $0.457 \pm 0.079$ & $-0.020 \pm 0.088$ \\
& 3 & $0.523 \pm 0.075$ & $0.004 \pm 0.088$ \\
& 4 & $0.508 \pm 0.076$ & $-0.020 \pm 0.088$ \\
\hline

\end{tabular}
\caption{Expectation values of stabilizer and destabilizer operators shown side by side for each Clifford operation shown in Fig.~\ref{fig : clifford_34}. The explicit form of the measured Pauli operators are listed in Table~\ref{tab:pauli_strings_narrow}.}
\label{tab : Clifford_34}
\end{table}

\section{Conclusions and Outlook}
\label{sec: Conclusions}

In this work, we introduced two complementary, scalable volumetric benchmarks, Clifford Volume (CLV) and Free-Fermion Volume (FFV), designed to quantify the computational capacity of quantum processors in the late-NISQ and early fault-tolerant regimes while remaining classically verifiable and platform independent. The core idea is to benchmark devices based on their ability to accurately implement randomly sampled unitaries from two practically and theoretically important operation classes: Clifford operations and free-fermion (fermionic linear optics) operations. Each class is efficiently simulable classically, enabling verification at scale, while together they form a universal gate set and serve as algorithmic primitives across diverse applications.

We formulated both benchmarks abstractly, without prescribing connectivity assumptions, native gate sets, or compilation strategies, ensuring that benchmark scores reflect achievable problem sizes rather than device-specific routing artifacts. We also provided concrete baseline realizations and numerical studies under a simple, transparent noise model to illustrate expected scaling behavior and demonstrate how the proposed threshold criteria respond to readout and entangling-gate errors. 

Finally, we experimentally validated the feasibility of CLV on real hardware by executing the protocol on Quantinuum's H2-1 trapped-ion device, determining a Clifford Volume of 34 under our experimental conditions. This demonstration confirms that the proposed framework can be implemented end-to-end on current systems and yields interpretable, reproducible device-level metrics.

Future directions include experimental FFV implementation across platforms to complement CLV results, and exploration of hybrid protocols combining both operation classes in single circuits. Since Clifford and free-fermion operations together form a universal gate set, such hybrid benchmarks, using circuits similar to those in Refs.~\cite{alam2025fermionic, Oszmaniec2022,bako2025fermionic}, could probe the boundary between classically simulable and quantum-advantage regimes. As systems progress toward fault tolerance, stricter fidelity-based metrics could replace current threshold-based approaches, providing continuity in this benchmarking methodology across the full spectrum from NISQ to fault-tolerant quantum computing.

\section*{Acknowledgments}

We thank Karl Mayer for valuable insights regarding the implementation of the benchmark protocols. We also thank Andr\'{a}s P\'{a}lyi and J\'{a}nos Asb\'{o}th  for many fruitful discussions. This work was supported by the Horizon Europe programme HORIZON-CL4-2022-QUANTUM-01-SGA via the project 101113946 OpenSuperQPlus100, and by the European Union’s Horizon Europe research and innovation program under Grant Agreement Number 101114305 (“MILLENION-SGA1” EU Project), and by the European Union’s Horizon Europe research and innovation program under grant agreement No 101135699 via the project SPINUS. We also acknowledge support by the Ministry of Culture and Innovation, and the National Research, Development and Innovation Office within the Quantum
Information National Laboratory of Hungary (Grant No. 2022-2.1.1-NL-2022-00004) and within  the National Research, Development and Innovation Fund, through the AI4QT project Nr. 2020-1.2.3-EUREKA-2022-00029 and Grant No. FK135220. A.P. also acknowledges support from the National Research, Development and Innovation Office of Hungary, Project No.\ C2245617 (KDP-2023).

\bibliographystyle{quantum}
\bibliography{references}

\newpage
\appendix

\noindent {\huge{\bf Appendices}}
\section{Step-by-step description of the CLV benchmark protocol}
\label{appendix : CV_step_by_step}

The Clifford Volume benchmark protocol shall start by determining  the candidate values (or range of values) of $n$, for which the CLV protocol shall be performed. For this purpose, one is allowed to use any appropriate strategy that is suitable for their resources. 
Possible strategies include applying a binary search, individual trials or numerical simulations using precise, device-specific error models.
For each possible value of $n$ the benchmark shall be evaluated by executing the steps outlined below:

\begin{enumerate}[leftmargin=2em]
\item \textit{Initialization:} 
\begin{enumerate}[label=(\alph*),leftmargin=9pt]
\item Set the value of $n$. 
\item Generate an ensemble of $K = 4$ randomly sampled $n$-qubit Clifford operations ($\mathcal{C}^k$, where $k \in \left[1,K \right]$), selected without replacement. 
\item For each operation in the ensemble, randomly select $4$ stabilizer generators $\mathcal{S}_i^k$ ($i \in [1,4]$) of the stabilizer group corresponding to the state prepared by the random Clifford operation. In addition, randomly select $4$ distinct $n$-qubit Pauli operators $\mathcal{D}_i^k$ ($i \in [1,4]$) that do not belong to the stabilizer group (destabilizers). 

\end{enumerate}

\item \textit{Circuit Preparation:} \vspace{1ex} \\
Construct a circuit implementation of each Clifford operation $\mathcal{C}^k$ in the ensemble. Compile each circuit according to the architecture of the quantum computer being benchmarked.

\item \textit{Circuit Execution:} \vspace{1ex} \\
\textit{\textbf{This step must be carried out on a quantum device.}}

For each circuit implementation, measure the expectation value of each corresponding Pauli operator $\mathcal{S}^m_i$ and $\mathcal{D}^m_i$ by the following:

\begin{itemize}[leftmargin=3pt]
\item[] For each $k$ from $1$ to $K=4$:
\begin{enumerate}[leftmargin=5pt]
\item[] For each $i$ from $1$ to $n_{\text{m}}=4$:
\begin{enumerate}[leftmargin=7pt]
\item[] For each $l$ from $1$ to $L \geq512$: 

\begin{itemize}[leftmargin=9pt]
\item[] Initialize the system in the $\left|0\right\rangle^{\otimes n}$ state, execute the quantum circuit corresponding to $\mathcal{C}^k$ and measure the expectation value $ \braket{\mathcal{S}_i^k} $.
\end{itemize}

\item[] For each $l$ from $1$ to $L \geq 512$:

\begin{itemize}[leftmargin=9pt]
\item[] Initialize the system in the $\left|0\right\rangle^{\otimes N}$ state, execute the quantum circuit corresponding to $\mathcal{C}^m$ and measure the expectation value $ \braket{\mathcal{D}_i^m} $ of the chosen non-stabilizer operator.
\end{itemize}
\end{enumerate}
\end{enumerate}
\end{itemize}

\item \textit{Performance Evaluation:} 

For the given $n$ the performance is considered successful if the following conditions are fulfilled simultaneously:

$$
\text{I.} \quad
\begin{cases}
\displaystyle \braket{\mathcal{S}_i^{k}}  - 2\sigma_{\mathcal{S}^k_i} \ge \tau_S, \\[3mm]
\displaystyle \left| \braket{\mathcal{D}_i^{k}} \right| + 2\sigma_{\mathcal{D}^k_i} \le \tau_D,
\end{cases}
\qquad \forall \, i,k
$$
where the standard deviation of the measured expectation value is $\sigma_{\mathcal{P}_i^{k}} = \sqrt{\bigl(1-\langle \mathcal{P}_i^{k}\rangle^2\bigr)/L}$.

$$
\text{II.} \quad
\begin{cases}
\displaystyle
\overline{\braket{\mathcal{S}^{\,k}}} - 5\,\overline{\sigma}_{\mathcal{S}^{\,k}} \ge \tau_S, \\[3mm]
\displaystyle
\left| \overline{\braket{\mathcal{D}^{\,k}}} \right| + 5\,\overline{\sigma}_{\mathcal{D}^{\,k}} \le \tau_D ,
\end{cases}
\qquad \forall\, k .
$$
where $\overline{\braket{\mathcal{S}^{\,k}}} $ and $\overline{\braket{\mathcal{D}^{\,k}}} $ denote the averages of the measured stabilizer and destabilizer expectation values corresponding to the $k$th Clifford unitary, and $\overline{\sigma}_{S^k}$ and $\overline{\sigma}_{D^k}$ are the associated standard deviations of these averages, defined as $\overline{\sigma}_{\mathcal{S}^k} =\frac{1}{4}\sqrt{\sum_{i=1}^{4} \sigma^{2}_{\mathcal{P}^{k}_{i}} }$. 
\end{enumerate}
Both the stabilizer and destabilizer conditions must be satisfied for all
measured observables and for every sampled Clifford unitary at a given qubit
count $n$.



\vspace{2ex}
\textit{Definition of the benchmark score:} The largest $n$ value for which the benchmark protocol succeeds.

\section{Step-by-step description of the FFV benchmark protocol}
\label{FFV_step_by_step}

The FFV benchmark procedure, similarly to the CLV benchmark, shall start by determining the candidate values for $n$. 
For each possible value of $n$ the benchmark shall be evaluated by executing the steps outlined below:

\begin{enumerate}[leftmargin=2em]
\item \textit{Initialization:} 
\begin{enumerate}[label=(\alph*),leftmargin=9pt]
\item Set the value of $n$. 
\item Generate an ensemble of $K = 4$ randomly sampled $O^m \in SO(2n)$ matrices, where $m \in \left[1, K \right]$. 
\item Determine the corresponding $n$-qubit free-fermion operations ($ F^m $) for each $O^k$. 
\begin{enumerate}[label=(\roman*),leftmargin=13pt]
    \item If $n\leq10$ select every element for each $O^k$, i.e., every Majorana operator to be measured in the following steps
    \item If $n > 10$, select the 
$$
n_{\text{m}} = 20 + \left\lfloor \frac{n}{20} \right\rfloor,
$$
largest elements by absolute value and store their indices ($I^k$) for each $O^k$, thereby restricting the number of Majorana operators to be measured in the following steps.  
\end{enumerate}
\end{enumerate}

\item \textit{Circuit Preparation:} 

Construct the circuit implementation of the free fermion operation in the ensemble. Compile each circuit according to the architecture of the quantum computer being benchmarked.

\item \textit{Circuit Execution:} 

\textbf{\textit{This step must be carried out on a quantum device.}}

For each free-fermion operation, measure the expectation value of each selected Majorana mode operator $\left\{m_i\right\}_{i \in I^k}$ by implementing the following steps:
\begin{itemize}[leftmargin=3pt]
\item[] For each $m$ from $1$ to $K=4$:
\begin{enumerate}[leftmargin=5pt]
\item[] For each $i$ from $1$ to $n_{\text{m}}$:
\begin{enumerate}[leftmargin=7pt]
\item[] For each $l$ from $1$ to $L\geq512$:
\vspace{1ex}
\begin{enumerate}[leftmargin=9pt]
 \item[] Initialize the system into a randomly chosen uniform superposition of fermionic states, execute the quantum circuit corresponding to $F^m$ and measure each $\braket{m_i}$ for $i \in I^k$, the expectation value of the Majorana mode operators.
\end{enumerate}
\end{enumerate}
\end{enumerate}
\end{itemize}

\item \textit{Performance Evaluation:} 


For the given $n$ the performance is considered successful if the following conditions are fulfilled simultaneously:
$$
\begin{cases}
\displaystyle
R_{ii}^{(J)}
- 2\,\sigma_{\parallel}^k
\;\ge\;
\dfrac{1}{e},
\\[3mm]
\displaystyle
\left|
R_{ij}^{(J)}
\right|
+ 2\,\sigma_{\perp}^k
\;\le\;
\dfrac{1}{2e},
\end{cases}
\,
\forall\, k ,
$$
where the statistical uncertainties $\sigma_{\parallel}^k$ and $\sigma_{\perp}^k$ are obtained by propagating the shot-noise uncertainties of the individual Majorana expectation values included in the linear combinations (Eq.~\eqref{eq : sigmas}), and the quantities $R_{ij}^{(J)}$ are defined in Eq.~\eqref{reduced_lincomb}.

\end{enumerate}

\textit{Definition of the benchmark score:} The largest $n$ value for which the benchmark protocol succeeds. 

\section{Circuit Compilation}
\label{appendix : circuit_compilation}

In this work, we deliberately do not prescribe any rule for how to compile or realize the Clifford or free-fermion operators on the hardware. 
Any implementation strategy ---including approximate or heuristic compilation methods---provided that the resulting circuit correctly implements the intended operator on the considered qubit subspace, may be used. 
In practice, experimental results shall be validated against the ideal expectation values computed from the exact operator.

From the perspective of evaluating the benchmark, the compiled circuit itself is not of intrinsic interest. 
Users may, for instance, employ ancilla qubits or intermediate encodings that make the executed circuit larger than the ideal logical operator. 
This has no impact on the benchmark score: only the \emph{dimension of the Clifford operator}, i.e., the number $n$, determines the benchmark score. Thus, implementing an $n$-qubit operator using $k$ additional ancilla qubits is still counted as an $n$-qubit realization of the benchmark. 
In what follows, we present representative compiled examples to illustrate how circuit resources scale with qubit count across different hardware platforms.

In the literature, there are extensive studies on the ideal circuit implementation of Clifford operators under different qubit coupling constraints. For all-to-all connectivity, the state-of-the-art theoretical upper bound on the two-qubit depth (defined as the number of parallel layers of entangling gates, with single-qubit operations assumed free) of an arbitrary $n$-qubit Clifford circuit is approximately $2n + O(\log^2 n)$ ~\cite{Maslov_2022}.
For linear–nearest-neighbor connectivity, a constructive synthesis achieves a depth of at most $7n-4$ by implementing Hadamard-free Clifford transformations in depth $5n$ and adding the required layers for full Clifford gates~\cite{maslov2023cnot}. In practice, however, current compilers and synthesis methods rarely reach these theoretical limits, instead, heuristic decompositions followed by local optimizations are typically employed.

\begin{figure}[t]
    \centering
    \includegraphics[width=0.9\linewidth]{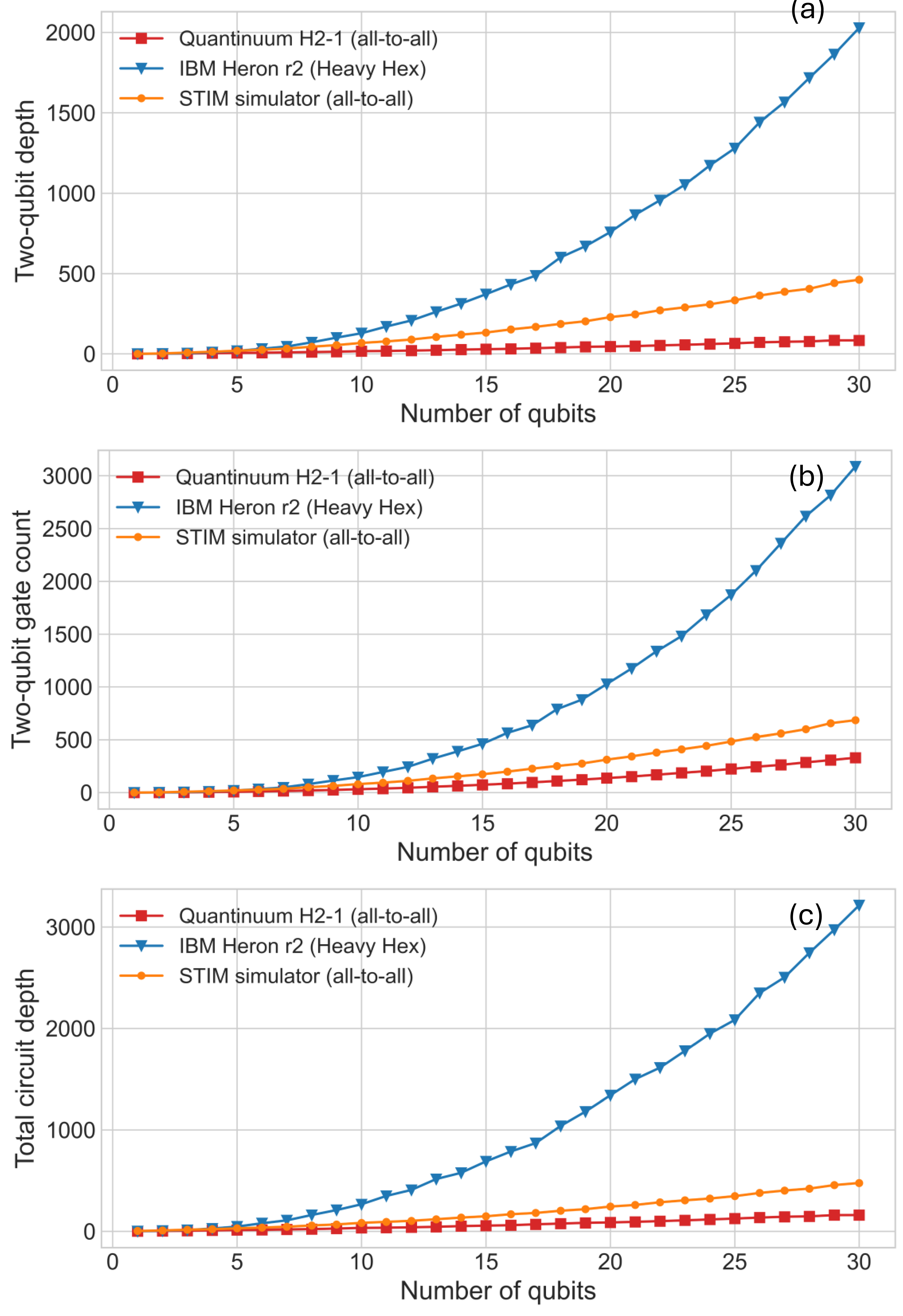}
    \caption{Characteristics of compiled circuit implementations of $n$-qubit Clifford operators across different quantum computational platforms. Each dataset compares the Quantinuum H2-1 (red squares), IBM Heron (blue triangles), and STIM simulator (orange circles) platforms. Subplots respectively show: (a) two-qubit circuit depth, (b) total two-qubit gate count, and (c) total circuit depth including single-qubit layers.}
    \label{fig: clifford_characteristics}
\end{figure}

We generated several example circuits according to the proposed benchmarks and compiled them for different platforms and connectivity topologies to illustrate the typical scaling of the compilation. Our goal was not to make a quantitative comparison between compilers or platforms, but rather to highlight the qualitative differences induced by more restricted topologies and by the use of different compilation strategies. The generated random circuits were compiled to native gate sets and connectivity constraints representative of three platforms: the Quantinuum H2-1 using TKET, the IBM Heron r2 using Qiskit, and a virtual all-to-all backend with single- and two-qubit Clifford gates as the native gate set, simulated with STIM. The figures show how key circuit metrics, such as the two-qubit depth (Fig.~\ref{fig: clifford_characteristics} (a)) , the two-qubit gate count (Fig.~\ref{fig: clifford_characteristics} (b)), and the total circuit depth (Fig.~\ref{fig: clifford_characteristics} (c)) scale with the number of logical qubits $n$. As expected, the connectivity topology dominates the scaling behavior. These results demonstrate that routing constraints primarily determine the resource overhead in compiled realizations (not the benchmark definition).

\begin{figure}[t]
    \centering
    \includegraphics[width=0.9\linewidth]{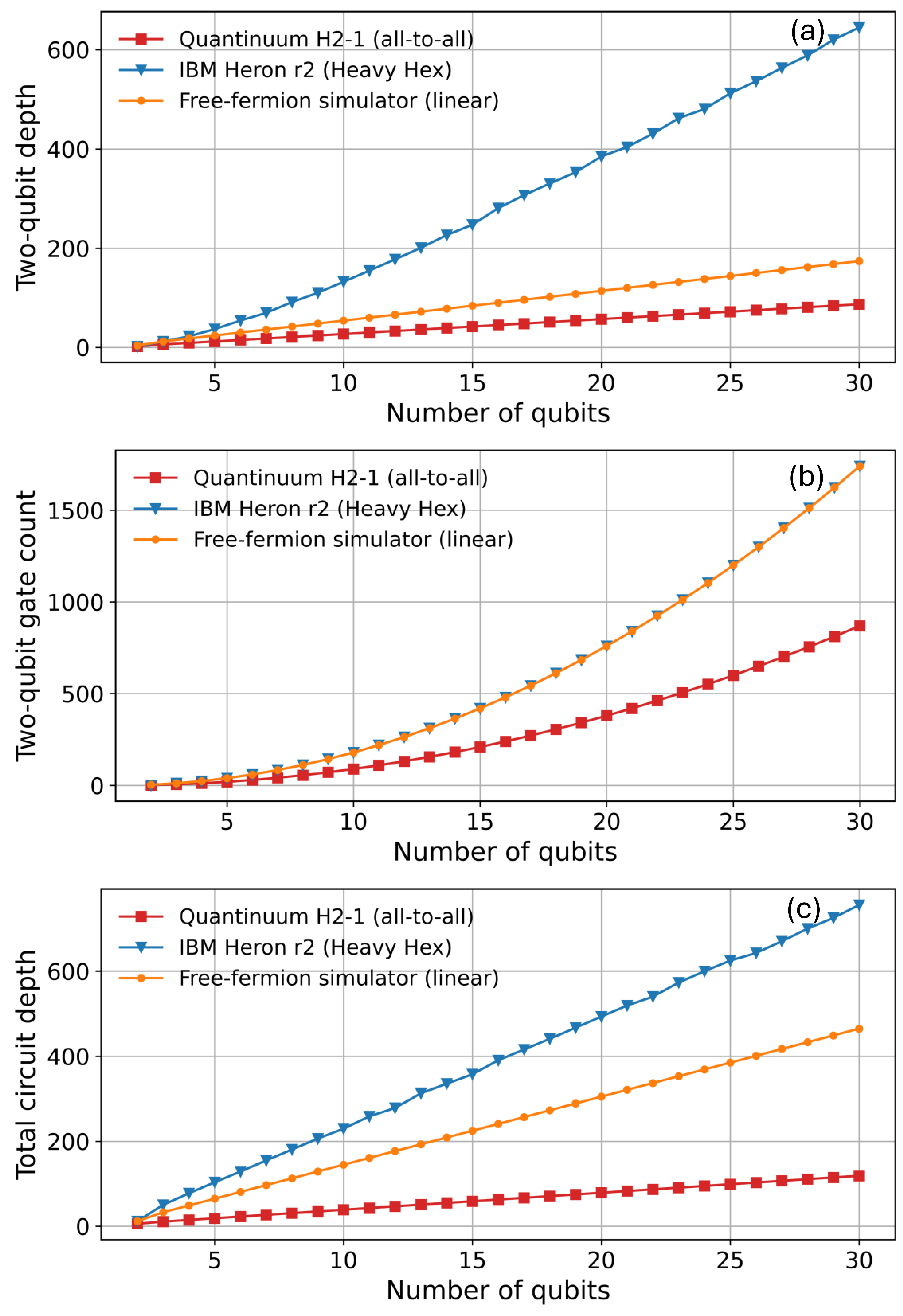}
    \caption{Characteristics of compiled circuit implementations of $n$-qubit free-fermion operators across different quantum computational platforms. Each dataset compares the Quantinuum H2-1 (red squares), IBM Heron (blue triangles), and our custom virtual backend based on Givens-rotation simulator (orange circles) platforms.
Subplots respectively show: (a) two-qubit circuit depth, (b) total two-qubit gate count, and (c) total circuit depth including single-qubit layers.
}
    \label{fig: freefermion_characteristics}
\end{figure}

In contrast to the case of the Clifford operators, there are few theoretical results establishing optimal or asymptotically minimal circuit depths for general $n$-qubit free-fermionic (matchgate) operators. Existing constructions are typically based on Givens-rotation or cosine-sine decompositions of the corresponding orthogonal transformation in $\mathrm{SO}(2n)$. Such factorizations require on the order of $n^2$ two-mode (i.e., two-qubit) rotations in total, and when restricted to nearest-neighbor couplings can be arranged in roughly $2n-1$ sequential layers. These values should be regarded as constructive upper bounds rather than proven optimal depths. To provide a practical perspective, we performed the same compilation analysis as for the Clifford benchmark, but using the free-fermionic circuits generated by the proposed procedure. The resulting circuits were compiled to the native gate sets and connectivity graphs of the IBM Heron r2 and Quantinuum H2-1 platforms, while for the ideal all-to-all case we employed our own simulator implementing the explicit Givens-rotation decomposition of random orthogonal transformations. The figures below summarize how the key circuit metrics two-qubit depth (Fig.~\ref{fig: freefermion_characteristics} (a)) , two-qubit gate count (Fig.~\ref{fig: freefermion_characteristics} (b)), and total circuit depth (Fig.~\ref{fig: freefermion_characteristics} (c)) scale with the number of logical qubits $n$ for these compiled free-fermion circuits.

Overall, the compiled results presented in Figs.~\ref{fig: clifford_characteristics} and~\ref{fig: freefermion_characteristics} exhibit clear and systematic scaling trends across different hardware topologies. For both Clifford and free-fermion operators, platforms with all-to-all connectivity show near-linear scaling of two-qubit circuit depth with qubit number, consistent with the theoretically expected behavior. In contrast, platforms with more constrained qubit coupling maps, such as those based on heavy-hex or similar topologies, display superlinear to approximately quadratic scaling ($(O(n^2))$) in both two-qubit depth and gate count, reflecting the dominant contribution of routing overheads required to satisfy limited qubit adjacency. The simulated, idealized backends (STIM and Givens-rotation simulators) approach the theoretical lower bounds on circuit depth, serving as baselines for optimal compilation under unrestricted connectivity. These results collectively emphasize that qubit connectivity and routing constraints—rather than intrinsic gate synthesis costs—are the primary determinants of resource scaling in compiled implementations of the proposed benchmarks.

\section{Reducing the Number of Measurements in the Free-Fermion Benchmark}
\label{appendix : reduce}

\begin{figure}[t]
    \centering
    \includegraphics[width=0.9\linewidth]{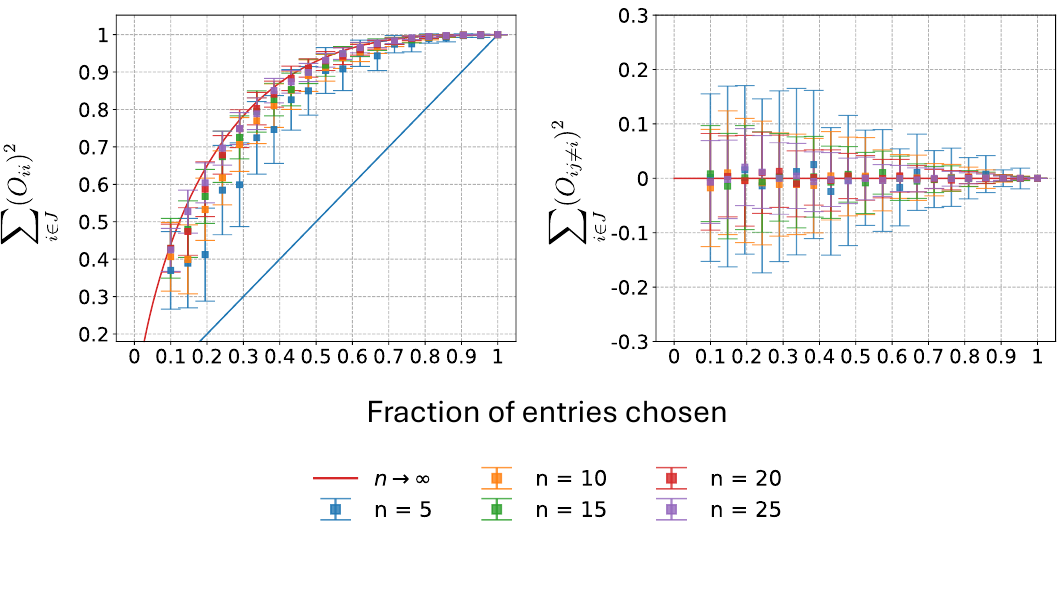}
    \caption{Statistical justification of the reducing strategy used in the free-fermion benchmark. The left panel shows the cumulative contribution to the sum of squares $\sum_{i \in S} (O_{ij})^2$ as a function of the fraction of entries selected (in descending order of absolute value), for various system sizes $n$. The solid blue line corresponds to a uniform (random) selection baseline. The right panel shows the deviation of the partial sums from the infinite-$n$ analytical approximation. }
    \label{fig : ratio}
\end{figure}

In what follows we justify the use of a cut-off in the number of measurements for the FFV benchmark by analyzing the statistical properties of entries in random orthogonal matrices. We will assume that, for the relevant number of qubits (where we apply the cut-off), the matrix elements can be approximated by a Gaussian distribution. Although entries of random special orthogonal matrices do not exactly follow a normal distribution -- due to the strong dependencies induced by orthogonality constraints -- we employ the common normal approximation. This approximation states that, for large matrices, the entries behave similarly to independent Gaussian random variables. Specifically, if $O \in SO(2n)$ is a $2n \times 2n$ special orthogonal matrix drawn from the uniform (Haar) measure, then each row of $O$ can be treated as having independent Gaussian entries with mean $0$ and variance $\frac{1}{2n}$. Indeed, $OO^T = I$ implies that each row of $O$ has unit norm. Geometrically we may view each row as a uniformly random vector on the $(2n-1)$-dimensional unit sphere. Since there is no preferred direction, it implies that
\begin{equation}
\overline{O_{ij}} = 0, \quad \text{and} \quad \overline{(O_{ij})^2} = \frac{1}{2n}.
\end{equation}
Let us introduce the notation $X = O_{ij} \sim N(0,\sigma^2)$ with $\sigma^2 = \frac{1}{2n}$. The corresponding  probability density function is then
\begin{equation}
f_X(x) = \frac{1}{\sqrt{2\pi}\,\sigma} \exp\!\left(-\frac{x^2}{2\sigma^2}\right).
\end{equation}
Let $\phi(x)$ and $\Phi(x)$ be the probability density function and cumulative distribution function of the standard normal distribution, respectively:
\begin{align}
\phi(x) & = \frac{1}{\sqrt{2\pi}} \exp\!\left(-\frac{x^2}{2}\right), \\\notag  \Phi(x) &= \int_{-\infty}^x \phi(t)\,dt.
\end{align}
We intend to determie the fraction $r$ of the $2n$ entries (in a given row) that exceed a threshold $T>0$ in absolute value. Since $X$ is symmetric,
\begin{align}
\Pr\left(|X| > T\right) &= 2\,\Pr(X > T) \\ \notag &= 2\left(1-\Phi\!\left(\frac{T}{\sigma}\right)\right) = r,
\end{align}
so that
\begin{equation}
T = \sigma\,\Phi^{-1}\!\left(1-\frac{r}{2}\right).
\end{equation}
Let us now compute the conditional expectation value of $X^2$ given $|X| > T$:
\begin{equation}
\braket{X^2 \,\Big|\, |X| > T} = \frac{2\int_{T}^{\infty} x^2 f_X(x)\,dx}{\Pr(|X|>T)}.
\end{equation}
If we substitute $x=\sigma u$, so that $dx=\sigma\,du$, and note that 
\begin{equation}
f_X(x) = \frac{1}{\sigma}\,\phi\!\left(\frac{x}{\sigma}\right),
\end{equation}
then we get
\begin{equation}
\int_{T}^{\infty} x^2 f_X(x)\,dx = \sigma^2 \int_{T/\sigma}^{\infty} u^2\,\phi(u)\,du.
\end{equation}
Thus,
\begin{equation}
\braket{X^2 \,\Big|\, |X| > T} = \frac{2\,\sigma^3 \int_{T/\sigma}^{\infty} u^2\,\phi(u)\,du}{r}.
\end{equation}
Let us observe that
\begin{equation}
\int_{T/\sigma}^{\infty} u^2\,\phi(u)\,du = \frac{T}{\sigma}\,\phi\!\left(\frac{T}{\sigma}\right) + \left(1-\Phi\!\left(\frac{T}{\sigma}\right)\right),
\end{equation}
then by using $1-\Phi(T/\sigma)=\frac{r}{2}$, we obtain
\begin{equation}
\lambda \equiv \braket{X^2 \,\Big|\, |X| > T} = \sigma^2 \left(1 + \frac{2T\,\phi(T/\sigma)}{\sigma\,r}\right).
\end{equation}
A row of $O$ has $2n$ entries, therefore selecting the top $r$ fraction in absolute value corresponds to $\lfloor r\,2n \rfloor$ entries. The expected contribution to the sum of squares from these entries is then
\begin{align}
&\left\langle\sum_{i\in J} (O_{ij})^2 \right\rangle \approx \\ 
&\left(r + 2\,\Phi^{-1}\!\left(1-\frac{r}{2}\right)\,\phi\!\left(\Phi^{-1}\!\left(1-\frac{r}{2}\right)\right)\right) \, ,
\end{align}
where, $J = \{\,i \,\mid\, |O_{ij}| > T(k)\}$ denotes the set of indices corresponding to the largest $|O_{ij}|$. For $r=10\%$, note that
\begin{equation}
\Phi^{-1}(0.95) \approx 1.645 \quad \text{and} \quad \phi(1.645) \approx 0.103.
\end{equation}
Hence,
\begin{equation}
\left\langle \sum_{i\in J} (O_{ij})^2 \right\rangle \approx 0.1 + 0.339 = 0.439,
\end{equation}
which is significantly larger than the value obtained by random selection. Our numerical findings show that, even for $n=10$ qubits, the expectation values obtained by selecting the top fraction of entries closely approximate the ideal ratio computed (see Figure \ref{fig : ratio}). Based on these results, we propose to use this reduced and renormalized quantity 
\begin{equation}
R_{ij}^{(J)} \equiv  \dfrac{1}{\lambda}\left\langle \sum_{k\in J} O_{ki} \braket{m_k}_{\rho_i} \right\rangle \approx 1
\end{equation}
for the benchmark, rather than employing the full linear combination, in order to reduce the resource demands of the benchmark.

\section{Pauli operators measured during the experiments.}
\label{appendix :paulis}

This appendix presents the Pauli operators whose expectation values were measured in the experiments, together with the corresponding measured expectation values. These operators correspond to the stabilizer and destabilizer observables associated with the Clifford unitaries discussed in the main text.

\begin{table}
\centering
\small
\setlength{\tabcolsep}{3pt}
\begin{tabular}{|c|c|}
\hline
\multicolumn{2}{|c|}{Stabilizers} \\ \hline \hline
$\mathcal{S}_1$ & {\scriptsize +IYYYXIXXYYIYYZXIXYIZYYZIYXXZXZ} \\ \hline
$\mathcal{S}_2$ & {\scriptsize +IIYZZZXXYXYIIZIYXXIZZYXYIYYYYI} \\ \hline
$\mathcal{S}_3$ & {\scriptsize +XXIIXIXIYYYXXYIIZXIYYXYXYYYZYI} \\ \hline
$\mathcal{S}_4$ & {\scriptsize -YIYZZZZZXZYXIXYZXZIYXIZZZZZZIY} \\ \hline \hline
\multicolumn{2}{|c|}{Destabilizers} \\ \hline \hline
$\mathcal{D}_1$ & {\scriptsize +XXIIXIXIYYYXXYIIZXIYYXYXYYYZYI} \\ \hline
$\mathcal{D}_2$ & {\scriptsize -ZXIZZZYZXIIXYXXIYYIZXXXYXXIZYI} \\ \hline
$\mathcal{D}_3$ & {\scriptsize +XIYZIYXXXZIIXYIXYIZIZIYXZYYYYX} \\ \hline
$\mathcal{D}_4$ & {\scriptsize -XXZIXIIXYIXZYXXXIIYXIYXXZXIZZZ} \\ \hline
\end{tabular}
\caption{Measured Pauli operators corresponding to Fig.~\ref{fig : clifford_30}, i.e., the $30$-qubit CLV benchmark implementation.}
\label{tab : pauli_strings_30}
\end{table}

\begin{table}
\centering
\small
\setlength{\tabcolsep}{3pt}
\begin{tabular}{|c|c|}
\hline
\multicolumn{2}{|c|}{Stabilizers} \\ \hline \hline
$\mathcal{S}_1$ & {\scriptsize +YXZIXXXXYXXYZIIXIYIXIYIIYIYYYYYIXYI} \\ \hline
$\mathcal{S}_2$ & {\scriptsize -YZZIXXYZZZZYYIXZYYYXIIYXIZXZZYYIXZY} \\ \hline
$\mathcal{S}_3$ & {\scriptsize +XXZIYYYXIZZIZIYZYZIZYXXIIZXYZYYXZZY} \\ \hline
$\mathcal{S}_4$ & {\scriptsize -XXXYXXZIZXZXZXZIZYZYYYXZZIXYZIXYXXI} \\ \hline \hline
\multicolumn{2}{|c|}{Destabilizers} \\ \hline \hline
$\mathcal{D}_1$ & {\scriptsize +YYZZIYZZZYZXYYXIZYXYYZXZYIZIXYXXZXX} \\ \hline
$\mathcal{D}_2$ & {\scriptsize -IIZYXZXYIYZIZXXXXYXZXIZZXIZZIZZXZXY} \\ \hline
$\mathcal{D}_3$ & {\scriptsize +XZIIZIXZZYZYXZIZIZXZXIXIZXXIYIIZZIY} \\ \hline
$\mathcal{D}_4$ & {\scriptsize +IYYZYZXYZYYZYIXXYXZXZXXXYXIIIZIZZIY} \\ \hline

\end{tabular}
\caption{Measured Pauli operators corresponding to Fig.~\ref{fig:clv_n35_n36} (a), i.e., the $35$-qubit CLV benchmark implementation.}
\label{tab : pauli_strings_35}
\end{table}

\begin{table}
\centering
\small
\setlength{\tabcolsep}{3pt}
\begin{tabular}{|c|c|}
\hline
\multicolumn{2}{|c|}{Stabilizers} \\ \hline \hline
$\mathcal{S}_1$ & {\scriptsize +XYZXXZZZYYYYYYYXIIXXYXZIXZXXZXYXZXZZ} \\ \hline
$\mathcal{S}_2$ & {\scriptsize -ZYIIXZXIXXIXYXIYIYXIZXXZYIYIIIZYYZYZ} \\ \hline
$\mathcal{S}_3$ & {\scriptsize +ZZXZXXXIYIXYZXIZYZZYXIXZZXZYYIIYIXXX} \\ \hline
$\mathcal{S}_4$ & {\scriptsize +XYXZIXXXYYXIYXIZYIIZZYYXIYIXZYXIYIZZ} \\ \hline  \hline

\multicolumn{2}{|c|}{Destabilizers} \\ \hline \hline
$\mathcal{D}_1$ & {\scriptsize +ZIIIXIZIZXZIZZIIZZZYYIYXYXIYYZXZIIZZ} \\ \hline
$\mathcal{D}_2$ & {\scriptsize -ZZIIXYZIZZYIYIYIZZYYIZIXXZIXZYYYXXIX} \\ \hline
$\mathcal{D}_3$ & {\scriptsize +ZYXIXXXXXXIYXXIXYZYIYZYZIYZZZIZZYZXI} \\ \hline
$\mathcal{D}_4$ & {\scriptsize +IXZXZYXYXZIIXYYZYZYYXIIXXIYXIXZXZIXX} \\ \hline

\end{tabular}
\caption{Measured Pauli operators corresponding to Fig.~\ref{fig:clv_n35_n36} (b), i.e., the $36$-qubit CLV benchmark implementation.}
\label{tab : pauli_strings_36}
\end{table}

\newpage


\begin{table}[t]
\centering
\small
\setlength{\tabcolsep}{3pt}
\begin{tabular}{|c|c|}
\hline
\multicolumn{2}{|c|}{Stabilizers} \\ \hline \hline

$\mathcal{S}^1_1$ & {\scriptsize +YZYYXIYYIYXZXIZYIXXYIXIYIZZXYYYYIY} \\ \hline
$\mathcal{S}^1_2$ & {\scriptsize +ZXYXIZZXYZZYYYYZXIIZZXYIIYIZYXXZYZ} \\ \hline
$\mathcal{S}^1_3$ & {\scriptsize -ZZIXXZZXIZIIXIYIYYXIXYZXYZZXIXXZZY} \\ \hline
$\mathcal{S}^1_4$ & {\scriptsize +YYIXIIXIYIZZIIIXZZXYIZXZZYXZIIXXIZ} \\ \hline

$\mathcal{S}^2_1$ & {\scriptsize +ZXYXIZYIYZYIYXIYZZYZZXXYZYYIZZIIZX} \\ \hline
$\mathcal{S}^2_2$ & {\scriptsize -ZIZZIYXIIIYXYYYIXIIYYZYZXZYIXYZYZY} \\ \hline
$\mathcal{S}^2_3$ & {\scriptsize +YZYXIIZXZXXXIIXZXXZZYIIZYYZYXYZYII} \\ \hline
$\mathcal{S}^2_4$ & {\scriptsize +YIIIZYZYZIYIIYXXIXXZZIYZYXXXXYIIXX} \\ \hline

$\mathcal{S}^3_1$ & {\scriptsize +ZXYXIZYIYZYIYXIYZZYZZXXYZYYIZZIIZX} \\ \hline
$\mathcal{S}^3_2$ & {\scriptsize -ZIZZIYXIIIYXYYYIXIIYYZYZXZYIXYZYZY} \\ \hline
$\mathcal{S}^3_3$ & {\scriptsize +YZYXIIZXZXXXIIXZXXZZYIIZYYZYXYZYII} \\ \hline
$\mathcal{S}^3_4$ & {\scriptsize +YIIIZYZYZIYIIYXXIXXZZIYZYXXXXYIIXX} \\ \hline\hline

$\mathcal{S}^4_1$ & {\scriptsize +ZXYXIZYIYZYIYXIYZZYZZXXYZYYIZZIIZX} \\ \hline
$\mathcal{S}^4_2$ & {\scriptsize -ZIZZIYXIIIYXYYYIXIIYYZYZXZYIXYZYZY} \\ \hline
$\mathcal{S}^4_3$ & {\scriptsize +YZYXIIZXZXXXIIXZXXZZYIIZYYZYXYZYII} \\ \hline
$\mathcal{S}^4_4$ & {\scriptsize +YIIIZYZYZIYIIYXXIXXZZIYZYXXXXYIIXX} \\ \hline\hline

\multicolumn{2}{|c|}{Destabilizers} \\ \hline\hline

$\mathcal{D}^1_1$ & {\scriptsize +IIXZIIIIXXZZYXZXZYXYYXIXYIZYZIIIZI} \\ \hline
$\mathcal{D}^1_2$ & {\scriptsize -YIZXYZIIYXIXZIZZYYYXZIXZXYYYIXZYXY} \\ \hline
$\mathcal{D}^1_3$ & {\scriptsize -YIYXXYXIZZZYXYYIXXYZYIXYYYZYIXXXXZ} \\ \hline
$\mathcal{D}^1_4$ & {\scriptsize +ZYIXXYYXIZZYXYYXZXIZYXXYIXIXZIYYIY} \\ \hline

$\mathcal{D}^2_1$ & {\scriptsize +XXZXZZIIZIIYXYIYXZXIIZXZXIIYXXYIYZ} \\ \hline
$\mathcal{D}^2_2$ & {\scriptsize +YIZIZYIZXIXXXYIZXIYYYXXZXYYZZIIXIY} \\ \hline
$\mathcal{D}^2_3$ & {\scriptsize +XIIXYYIXYZIZIXXYYYIIXYXZYYZXIYZXZX} \\ \hline
$\mathcal{D}^2_4$ & {\scriptsize -XIZZIYIYZIZYXYYZIIZZIIZZZIYXXXZZXY} \\ \hline

$\mathcal{D}^3_1$ & {\scriptsize +XXZXZZIIZIIYXYIYXZXIIZXZXIIYXXYIYZ} \\ \hline
$\mathcal{D}^3_2$ & {\scriptsize +YIZIZYIZXIXXXYIZXIYYYXXZXYYZZIIXIY} \\ \hline
$\mathcal{D}^3_3$ & {\scriptsize +XIIXYYIXYZIZIXXYYYIIXYXZYYZXIYZXZX} \\ \hline
$\mathcal{D}^3_4$ & {\scriptsize -XIZZIYIYZIZYXYYZIIZZIIZZZIYXXXZZXY} \\ \hline

$\mathcal{D}^4_1$ & {\scriptsize +XXZXZZIIZIIYXYIYXZXIIZXZXIIYXXYIYZ} \\ \hline
$\mathcal{D}^4_2$ & {\scriptsize +YIZIZYIZXIXXXYIZXIYYYXXZXYYZZIIXIY} \\ \hline
$\mathcal{D}^4_3$ & {\scriptsize +XIIXYYIXYZIZIXXYYYIIXYXZYYZXIYZXZX} \\ \hline
$\mathcal{D}^4_4$ & {\scriptsize -XIZZIYIYZIZYXYYZIIZZIIZZZIYXXXZZXY} \\ \hline

\end{tabular}
\caption{Measured Pauli operators corresponding to Fig.~\ref{fig : clifford_34}, i.e., the $34$-qubit CLV benchmark implementation.}
\label{tab:pauli_strings_narrow}
\end{table}

\end{document}